\documentclass[aps,amssymb,amsmath,pra,twocolumn,showpacs,superscriptaddress]{revtex4-1}

\usepackage{graphicx}
\usepackage{dcolumn}
\usepackage{bm}
\usepackage{subfigure}
\usepackage{color}

\usepackage{amssymb,amsmath,amsfonts,latexsym,graphicx,verbatim}

\usepackage[margin=0.6in]{geometry}

\usepackage[english]{babel}
\usepackage{times}
\usepackage{latexsym}
\usepackage{fancyhdr}
\usepackage{float}
\usepackage{afterpage}
\usepackage{dsfont}

\usepackage{multirow}

\usepackage{upgreek}

\definecolor{Ablue}{rgb}{0.96,0.24,0.00}

\definecolor{Abluetitle}{rgb}{0.,0.24,0.51}
\newcommand{\bluetitle}{\color{Abluetitle}}

\definecolor{orange}{rgb}{0.96,0.24,0.00}

\definecolor{darkred}{rgb}{0.55, 0.0, 0.0}

\definecolor{Gray}{gray}{0.85}
\definecolor{LightCyan}{rgb}{0.88,1,1}
\definecolor{darksalmon}{rgb}{0.91, 0.59, 0.48}
\definecolor{maroon}{cmyk}{0,0.87,0.68,0.32}

\definecolor{mustard}{rgb}{1.0, 0.86, 0.35}
\newcolumntype{a}{>{\columncolor{Gray}}c}
\newcolumntype{b}{>{\columncolor{white}}c}

\usepackage{array}
\newcolumntype{L}[1]{>{\raggedright\let\newline\\\arraybackslash\hspace{0pt}}m{#1}}
\newcolumntype{C}[1]{>{\centering\let\newline\\\arraybackslash\hspace{0pt}}m{#1}}
\newcolumntype{R}[1]{>{\raggedleft\let\newline\\\arraybackslash\hspace{0pt}}m{#1}}

\usepackage[colorlinks=true , citecolor=blue,urlcolor=blue]{hyperref}

\newcommand{\id}{\mathds{1}}

\newcommand{\vxe}{\varepsilon}

\newcommand{\xg}{\gamma}
\newcommand{\xt}{\theta}

\newcommand{\xl}{\lambda}
\newcommand{\xr}{\rho}
\newcommand{\xo}{\omega}

\newcommand{\app}{\approx}

\newcommand{\trepol}{t_{\R{repol}}}
\newcommand{\Bpol}{\textbf{B}_{\R{pol}}}

\newcommand{\Bp}{B_{\R{pol}}}
\newcommand{\Br}{B_{\R{relax}}}
\newcommand{\Cs}{{}^{13}\R{C}}

\newcommand{\Hs}{{}^{1}\R{H}}
\newcommand{\Sis}{{}^{29}\R{Si}}

\newcommand{\mB}[0]{\mathcal{B}}
\newcommand{\dxo}[0]{\dot\omega}

\newcommand{\xD}{\Delta}

\newcommand{\xL}{\Lambda}
\newcommand{\xO}{\Omega}

\newcommand{\bz}[0]{\hat{\mathbf z}}

\newcommand{\fr}[2]{\frac{#1}{#2}}

\newcommand{\wt}[1]{\widetilde{#1}}

\newcommand{\mH}[0]{\mathcal{H}}

\newcommand{\beq}{\begin{equation}}
\newcommand{\eeq}{\end{equation}}
                  
\newcommand{\benum}{\begin{enumerate}}
\newcommand{\eenum}{\end{enumerate}}
                    
\newcommand{\bit}{\begin{itemize}}
\newcommand{\eit}{\end{itemize}}

\newcommand{\bea}{\begin{eqnarray}}
\newcommand{\eea}{\end{eqnarray}}

\newcommand{\bev}{\begin{verbatim}}
\newcommand{\eev}{\end{verbatim}}

\newcommand{\zt}{\times}

\newcommand{\B}[1]{\mathbf{#1}}
\newcommand{\T}[1]{\textbf{#1}}
\newcommand{\I}[1]{\textit{#1}}
\newcommand{\R}[1]{\textrm{#1}}

\newcommand{\zfl}[1]{\protect\label{fig:#1}}
\newcommand{\zfr}[1]{Fig. \ref{fig:#1}}

\newcommand{\ket}[1]{\left\vert{#1}\right\rangle}

\newcommand{\sand}[3]{\langle{#1}\vert#2\vert{#3}\rangle}

\newcommand{\ba}{\left\{ \begin{array}{lr}}
\newcommand{\ea}{\end{array}\right.}

\definecolor{darkred}{rgb}{0.55, 0.0, 0.0}

\newcommand{\blist}[1]{
 \begin{list}{#1}
 \begin{align}
	 arrow
 \end{align}
 $\checkmark\star
  { \setlength{\itemsep}{3pt}
     \setlength{\parsep}{2pt}
     \setlength{\topsep}{3pt}
     \setlength{\partopsep}{0pt}
     \setlength{\leftmargin}{1em}
     \setlength{\labelwidth}{1em}
     \setlength{\labelsep}{0.5em} } }
\newcommand{\elist}{
  \end{list}  }

\DeclareMathSymbol{\vartheta}{\mathalpha}{letters}{"12}
\DeclareMathSymbol{\theta}{\mathalpha}{letters}{"23}
\DeclareMathSymbol{\phi}{\mathalpha}{letters}{"27}
\DeclareMathSymbol{\varphi}{\mathalpha}{letters}{"1E}

\newcommand{\bef}
{
\begin{figure}[htbp]
\centering
}

\newcommand{\eef}{\end{figure}}

\newcommand{\beginsupplement}{%
        \setcounter{table}{0}
        \renewcommand{\thetable}{S\arabic{table}}%
        \setcounter{figure}{0}
        \renewcommand{\thefigure}{S\arabic{figure}}%
     }

\newcommand{\affA}{Department of Chemistry, University of California Berkeley, and Materials Science Division Lawrence Berkeley National Laboratory, Berkeley, California 94720, USA.}
\newcommand{\affB}{Department of Physics, and CUNY-Graduate Center, CUNY-City College of New York, New York, NY 10031, USA.}
\newcommand{\affD}{Department of Chemical and Biomolecular Engineering, and Materials Science Division Lawrence Berkeley National Laboratory University of California, Berkeley, California 94720, USA.}
\newcommand{\affE}{Fakultat Physik, Technische Universitat Dortmund, D-44221 Dortmund, Germany.}

\newcommand{\affG}{Nuclear Magnetic Resonance Facility, University of California Davis, Davis, California 95616, USA.}

\begin{document}
\title{\bluetitle{Room temperature ``\I{Optical Nanodiamond Hyperpolarizer}'': physics, design and operation}}


  \author{A. Ajoy}\email{ashokaj@berkeley.edu}\affiliation{\affA}
  \author{R. Nazaryan}\affiliation{\affA}
  \author{E. Druga}\affiliation{\affA}
  \author{K. Liu}\affiliation{\affA}
 \author{A. Aguilar}\affiliation{\affA}
  \author{B. Han}\affiliation{\affA}
  \author{M. Gierth}\affiliation{\affA}
  \author{J. T. Oon}\affiliation{\affA}
  \author{\\B. Safvati}\affiliation{\affA}
  \author{R. Tsang}\affiliation{\affA}
  \author{J. H. Walton}\affiliation{\affG}
    \author{D. Suter}\affiliation{\affE}
  \author{C. A. Meriles}\affiliation{\affB}
  \author{J. A. Reimer}\affiliation{\affD}
  \author{A. Pines}\affiliation{\affA}

\begin{abstract}
Dynamic Nuclear Polarization (DNP) is a powerful suite of techniques that deliver multifold signal enhancements in NMR and MRI. The generated \I{athermal} spin states can also be exploited for quantum sensing and as probes for many-body physics. Typical DNP methods require use of cryogens, large magnetic fields, and high power microwaves, which are expensive and unwieldy. Nanodiamond particles, rich in Nitrogen-Vacancy (NV) centers, have attracted attention as alternative DNP agents because they can potentially be optically hyperpolarized at room temperature. Indeed the realization of a miniature \I{``optical nanodiamond hyperpolarizer''}, where $\Cs$ nuclei are optically hyperpolarized has been a longstanding goal but has been technically challenging to achieve. Here, unravelling new physics underlying an optical DNP mechanism first introduced in [Ajoy \I{et al.}, Sci. Adv. 4, eaar5492 (2018)], we report the realization of such a device in an ultracompact footprint and working fully at room temperature. Instrumental requirements are \I{very} modest: low polarizing fields, extremely low optical and microwave irradiation powers, and convenient frequency ranges that enable device miniaturization. We obtain best reported optical $\Cs$ hyperpolarization in diamond particles exceeding 720 times of the thermal 7T value (0.86\% bulk polarization), corresponding to a ten-million-fold gain in NMR averaging time. In addition the hyperpolarization signal can be \I{background-suppressed} by over two-orders of magnitude and retained for multiple-minute long periods. Besides compelling applications in quantum sensing, and bright-contrast MRI imaging, this work paves the way for low-cost DNP platforms that relay the $\Cs$ polarization to liquids in contact with the high surface-area particles. This will ultimately allow development of miniature ``quantum-assisted'' NMR spectrometers for chemical analysis.

\end{abstract}

\maketitle

\T{\I{Introduction:}} -- Dynamic Nuclear Polarization (DNP)~\cite{Carver53}, the process of transferring spin polarization from electrons to surrounding nuclei, \I{hyperpolarizing} them to levels of large fictitious magnetic fields~\cite{Abragam78}, has been a burgeoning field with a multitude of applications across numerous disciplines. For instance the versatile, noninvasive and chemical specific spectroscopic and imaging techniques~\cite{Ernst} of NMR and MRI can see their signals vastly enhanced by several orders of magnitude through the use of DNP~\cite{Ardenkjaer15}. Moreover hybrid electron-nuclear quantum sensing platforms, for instance gyroscopes~\cite{Ajoy12g,Donley10}, can see large sensitivity gains through nuclear hyperpolarization. Finally the generated \I{athermal} spin states also provide valuable physical testbeds to study many body quantum dynamics, spin transport and localization in dipolar coupled spin systems~\cite{Goldman,deluca15}. That said however, conventional methods of DNP~\cite{ArdenkjaerLarsen03,Maly08,Rosay10} involve the use of cryogenic conditions ($\lesssim$1K) and high magnetic fields ($\gtrsim$3T) in order to first  generate the electron polarization. This often entails a limited hyperpolarization throughput given several hour-long relaxation times and the necessity to cool down the sample every time. In many respects, this restricts wider real-world applications given the steep cost (often $>\$$1M) of setup and maintenance of cryogenic polarizer devices. There is, therefore, a strong desire for inexpensive \I{room-temperature} DNP platforms that can potentially retrofit existing NMR/MRI infrastructure, and provide hyperpolarization generation \I{``at source''}~\cite{Ji17}.

In recent years, the development of new classes of ``quantum materials'' in wide bandgap semiconductors has proffered them as exciting candidates for optical room temperature hyperpolarization. This leverages the fact that electronic spin defects in the systems can be optically initialized~\cite{Jelezko06}. Indeed \I{nanodiamond} particles rich in Nitrogen Vacancy (NV) defect centers have been suggested as compelling new hyperpolarization platforms~\cite{Fischer13,London13,,Abrams14}. The NV electrons can be optically polarized with modest resources, the polarization transferred to $\Cs$ nuclei in the lattice, and subsequently relayed to nuclei in a liquid in contact with the high surface-area ($\gtrsim$10m$^2$/g) particle surfaces~\cite{Ajoy17}. Hyperpolarized particulate diamonds come with a plethora of other applications: powdered samples represent the optimal size configuration to fill a sensor volume and maximize the number of spins available for quantum sensing~\cite{Ajoy12g}. There are also compelling possibilities for quantum sensors constructed from single levitating hyperpolarized diamond particles~\cite{Hoang16}. Finally, nanodiamonds are fluorescent and non-toxic, and hyperpolarization opens new pathways for dual-mode optical and MRI imaging of particles targeting disease locations~\cite{Schrand07,yu05,Chang08}.

In recent work we had demonstrated the first experimental technique for the generation of optically hyperpolarized diamond particles~\cite{Ajoy17}. We relied on an unconventional DNP regime operational at low fields and demonstrated hyperpolarization of the $\Cs$ nuclei irrespective of the orientation of the individual crystallites. In this paper, we refine the experimental conditions for the optimal generation of hyperpolarization, by focusing on the field dependence of the underlying DNP process. We conclusively demonstrate that this optical DNP mechanism is \I{fully} orientation independent and \I{low-field only}, occurring at polarization fields $\Bp$=1-70mT, and low-enough to be generated with the simplicity of a refrigerator magnet. Besides magnetic field, we demonstrate that the DNP mechanism affords extremely benign requirements for optical and microwave (MW) excitation. 

Leveraging this new physics, we construct a room temperature \I{``optical nanodiamond hyperpolarizer''} device that can produce $\Cs$ hyperpolarization in diamond particles with high throughput ($\app$20mg/min). \zfr{prototype}A shows a photograph and rendered view of the device.  Ref. ~\cite{hypercubevideo} shows a video of the device in operation. It has an ultracompact form-factor ($\sim$10in. edge) that houses all the electronic, MW and optical components required to produce hyperpolarization in diamond particles at room temperature.  The device is portable enough to ``\I{DNP-retrofit}'' any magnet system. We have deployed three such devices across 7T, 9.4T and 7T (imaging) magnets at UC Berkeley and UC Davis.  We emphasize that the ability to construct such a hyperpolarization device is \I{itself} very surprising and nontrivial, and due to a remarkable confluence of factors unique to the physics of optically pumped quantum defects. In this paper we uncover these features and also highlight engineering design aspects that make possible the miniaturizable device.

\T{\I{Hyperpolarization results:}} -- We begin in \zfr{prototype}B-E by demonstrating representative results using our hyperpolarizer device on various types of diamond samples. The $\Cs$ hyperpolarization enhancements $\varepsilon$ are evaluated with respect to the thermal Boltzmann level at 7T, corresponding to a time acceleration for spectroscopy or imaging $t_{\R{acc}}\app\varepsilon^2 \fr{T_1(7T)}{T_1(\Bp)}$. The key advantage of optical hyperpolarization is that the enhancement factor is not theoretically bounded, unlike in conventional methods wherein $\vxe\leq \xg_e/\xg_n$, the ratio of the electronic and nuclear gyromagnetic ratios. For a typical single crystal sample (4$\zt$4$\zt$0.25mm) we obtain in \zfr{prototype}B large DNP enhancements $\varepsilon=$950 corresponding to a $\Cs$ polarization level $\sim$1.1\% and larger than \I{ten-millon} fold acceleration ($t_{\R{acc}}\app 5.3\zt 10^7$) in averaging time. For randomly oriented microcystalline diamond powder (see \zfr{prototype}B), we obtain the best reported polarization, with $\varepsilon=$720, corresponding to a $\Cs$ polarization level $\sim$0.86\%, and an acceleration factor $t_{\R{acc}}\app 9.8\zt 10^6$. We note that while cyrogenic DNP provides larger enhancements, a vast majority of the obtained hyperpolarization is lost upon sample transfer outside the cryostat~\cite{Rej15,Bretschneider16}. The polarization in \zfr{prototype}C is thus ultimately stronger, obtained at higher throughput, and optically replensible. Moreover, due to the DNP being carried out under ambient conditions, the hyperpolarization enhancements can be maintained even when the particles are immersed in solution: for instance, common solvents like water, DMSO, oil, and biologically relative liquids such as saline and blood (see Supplementary Information~\cite{SOM}). The microcrystals in \zfr{prototype}C are employed for all the other experiments reported in this manuscript. 

Given the high polarization gain (see \zfr{prototype}D), we are also able to detect a \I{single} hyperpolarized diamond microparticle with a \I{single-shot} SNR $>$40 (see \zfr{prototype}D(\I{ii})). This is remarkable because the particle occupies a very small part of the detection coil ($\sim$1cm. cylinder) we employ, with sample fill-factor for NMR detection $\app 6\zt10^{-6}$. The high signals pave the way for several quantum sensing applications constructed out of hybrid spin-mechanical systems in single microparticles~\cite{Wan16,Ma17}.   Finally \zfr{prototype}E demonstrates results with the device on a \I{large} mass $\app$42mg of commercially available 100nm nanodiamond particles in solution (see \I{Materials}). We ascribe the lower polarization herein to be limited by material properties and finite optical penetration through the colloidal suspension. In a forthcoming manuscript, we study these material conditions in detail, and demonstrate methods by which the diamond particles can be rendered more suitable for optical DNP.

\begin{figure}[!htp]
  \centering
	{\includegraphics[width=0.48\textwidth]{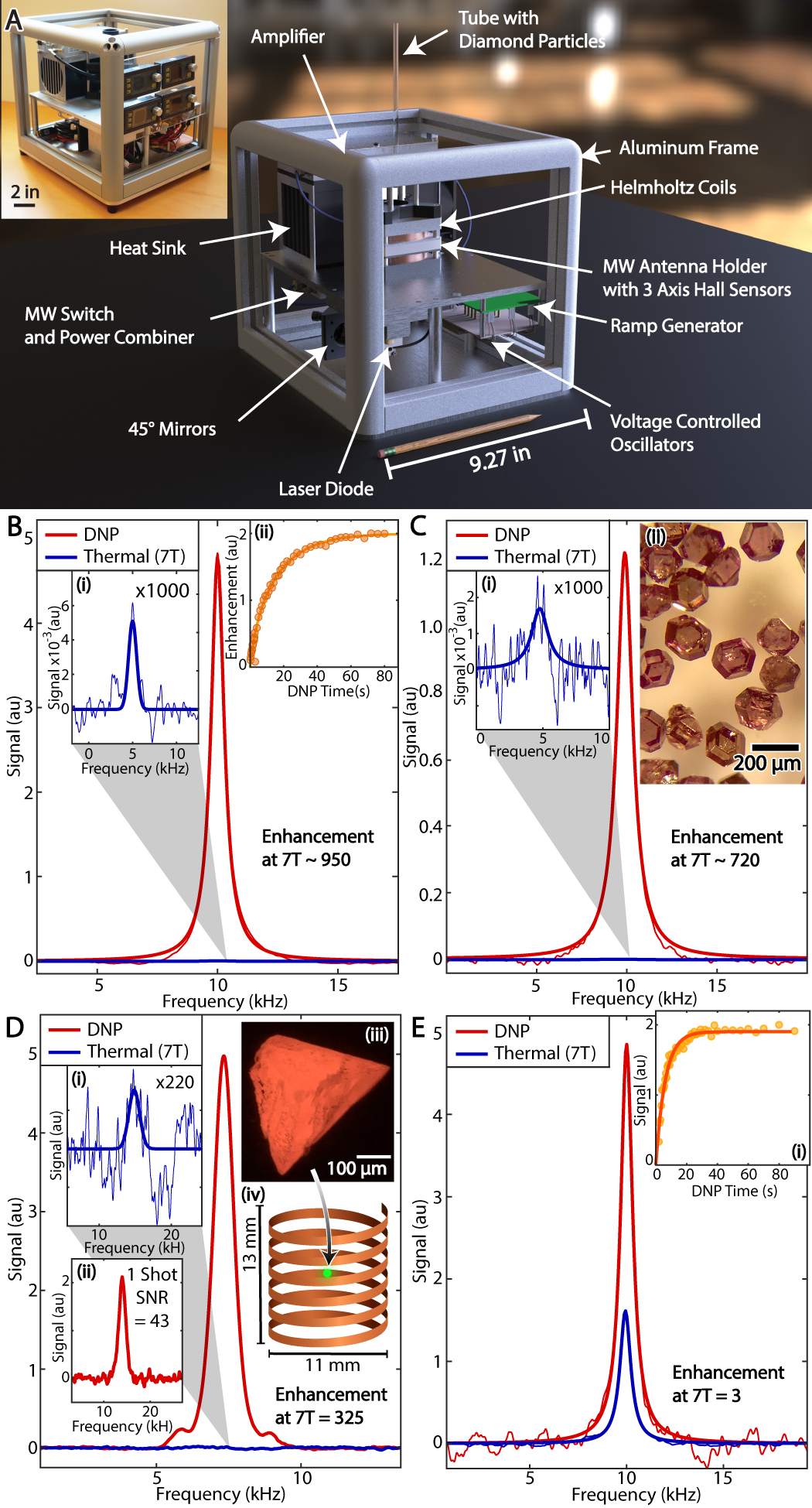}}
  \caption{\T{Nanodiamond hyperpolarizer.} (A) \I{Rendered model} on a table-top. Compact device ($<$10in.cubical edge) containing all optical, microwave (MW), and control components generates hyperpolarized diamond particles at high throughput $\app$20mg/min. \I{Inset:} Photograph of the constructed device. See Ref.~\cite{hypercubevideo} for video of operation. (B-D) \I{Results using hyerpolarizer device} at 38mT. $\Cs$ signals are measured by transfer to a 7T NMR magnet. Red (blue) lines show the DNP signal (7T thermal signal, zoomed in insets).  (B) \I{Single crystal} DNP, demonstrating enhancement $\varepsilon\app$950. Hyperpolarization buildup, typical for most samples, occurs in under 60s. (C) \I{Microcystalline diamond powder} DNP, with best reported polarization enhancements $\varepsilon\app$720. \I{Inset:} Particle micrograph. (D) \I{Single particle}  DNP of a 10\% $\Cs$ enriched and $\sim$400$\mu$m sized single particle showing $\varepsilon=$325 over 7T.  Inset \I{(ii)}: Single shot DNP signal with SNR $\app$43 in a 1235mm$^{3}$ NMR coil. Inset \I{(iii)}: Particle micrograph with fluorescence from NV centers. Inset \I{(iv)}: Schematic of detection coil. We obtain large single shot signals despite poor sample fill-factors $\sim 10^{-6}$. (C) \I{Nanodiamond} DNP with commercial 100nm NDs in solution, showing $\varepsilon=$3 over 7T.  \I{Inset:} Hyperpolarization buildup curve.}
 \zfl{prototype}	
\end{figure}

\T{\I{Technology enabling miniaturization:}} -- Hyperpolarization is achieved by the continuous application of laser and frequency swept MW irradiation at low background fields $\Bpol$~\cite{Ajoy17} (see \zfr{theory}B). The laser polarizes NV centers (we estimate to $>$10\%) to the $m_s$=0 sublevel, and the MWs transfer the polarization to the $\Cs$ nuclei. In practice we use a combination of three cascaded MW sweepers for greater signal enhancement~\cite{Ajoy18}. Refining and expanding on the work in Ref.~\cite{Ajoy17}, we begin by first summarizing the key results of this paper. Miniaturized hyperpolarization devices are at all possible because of a surprising confluence of four factors stemming from counterintuitive attributes of the underlying DNP mechanism:

\I{(i)} \I{Field:} -- We observe that hyperpolarization is optimum at a low polarizing field $\Bpol\app$38mT. Such low fields are simple to produce in a the miniature footprint either through a permanent magnet or a simple magnetic coil. There are absolutely no requirements on field alignment. Given that we are dealing with powders with inherently broadened electronic spectra, there are no constraints on field homogeneity, in principle DNP works even with inhomogeneities $\xD\Bp\sim \Bp$. Moreover, due to spin reorientation during sample shuttling, the polarization field does not even need to be aligned with the detection NMR field. This allows installation and retrofits of the device anywhere in the vicinity of detection magnets.  

\I{(ii)} \I{Optics:} -- the laser excitation required is of very low power ($\app$ 33mW/$\R{mm}^3$),  since the NVs need to be polarized only once every $T_{1e}$ relaxation cycle. We contrast this with much higher optical powers $\sim 1$mW/$(\mu$m$)^2$ required for conventional quantum sensing experiments employing optical NV center readout. Most importantly, there are no requirements on excitation wavelength (510nm$\lesssim\xl\lesssim$575nm), linewidth, or mode quality. Indeed diffuse irradiation through multimode optical fibers are sufficient. The optical excitation is in completely \I{cw}-mode, requiring no synchronization or pulsing infrastructure (eg. AOMs). This facilitates the use of inexpensive miniaturizable laser diode excitation sources.

\I{(iii)} \I{Microwaves:} -- the MW power is also exceedingly low ($\app$ 2mW/$\R{mm}^3$). There is a relatively weak dependence of the DNP enhancement on MW power~\cite{Ajoy17}, making the hypolarization robust to MW inhomogeneity. We have measured the electron Rabi frequency $\xO_e\app$430kHz and a MW inhomogeneity of $\app$12\% from the 4mm loop antenna in our device~\cite{SOM}. Moreover, since the MWs are chirped, they are inherently immune to carrier phase noise. Our MW excitation linewidth, for instance, is $\app$10MHz. The frequency sweep band ($\mB\sim$3.64-4GHz) at $\Bp$=38mT lies in the commercial WiMAX regime, and the use of commercial chip scale voltage controlled oscillator (VCO) sources ubiquitously available and this band greatly simplify miniaturization. Broadband antennas can deliver the MWs, with no requirement for a MW cavity or sophisticated transmission infrastructure. MW chirp repetition rates are slow $\xo_r\app$147Hz, robust $\xD\xo_r\app$53Hz, and to a good approximation \I{independent} of polarizing magnetic field. This allows simple frequency sweep infrastructure and the cascading of multiple MW sources to boost DNP efficiency~\cite{Ajoy18}.

\I{(iv)} \I{Polarization sign:} -- Sweeps over every part of the NV electronic spectrum produces hyperpolarization that constructively adds to the \I{same} polarization sign.  The MW sweep bands can be optimally tuned to the simulated electronic spectral widths at any given $\Bp$ field to optimize final hyperpolarization enhancements.

These attributes are consequences of the underlying DNP mechanism. We refer the reader to more detailed expositions elsewhere~\cite{Zangara18}, but briefly mention that DNP occurs when the nuclear Larmor frequency $\xo_L=\xg_n\Bp\app$ 10-700kHz is smaller than the hyperfine coupling $A$, i.e. $\xo_L<|A|$. Here $\xg_n$ is the $\Cs$ gyromagnetic ratio, $\xg_n\app$10.7kHz/mT. The swept MWs excite a \I{sequential} set of Landau-Zener (LZ) crossings between the electron-nuclear spin states in the rotating frame, and this drives a \I{``ratchet''} type process for polarization transfer.  This immediately means that both the laser and microwave powers are necessarily low: MW powers low enough to maintain adiabiaticity of the LZ traversals, and  the laser excitation sufficiently low power to not break the coherence of the polarization transfer process. Indeed at optical powers that we operate under, the NV electronic repolarization occurs at a rate $\app 1/T_{1e}$, and takes place predominantly during the long intervals far away from the LZ anti-crossings~\cite{Ajoy17,Zangara18}.

\begin{figure}[t]
  \centering
	{\includegraphics[width=0.5\textwidth]{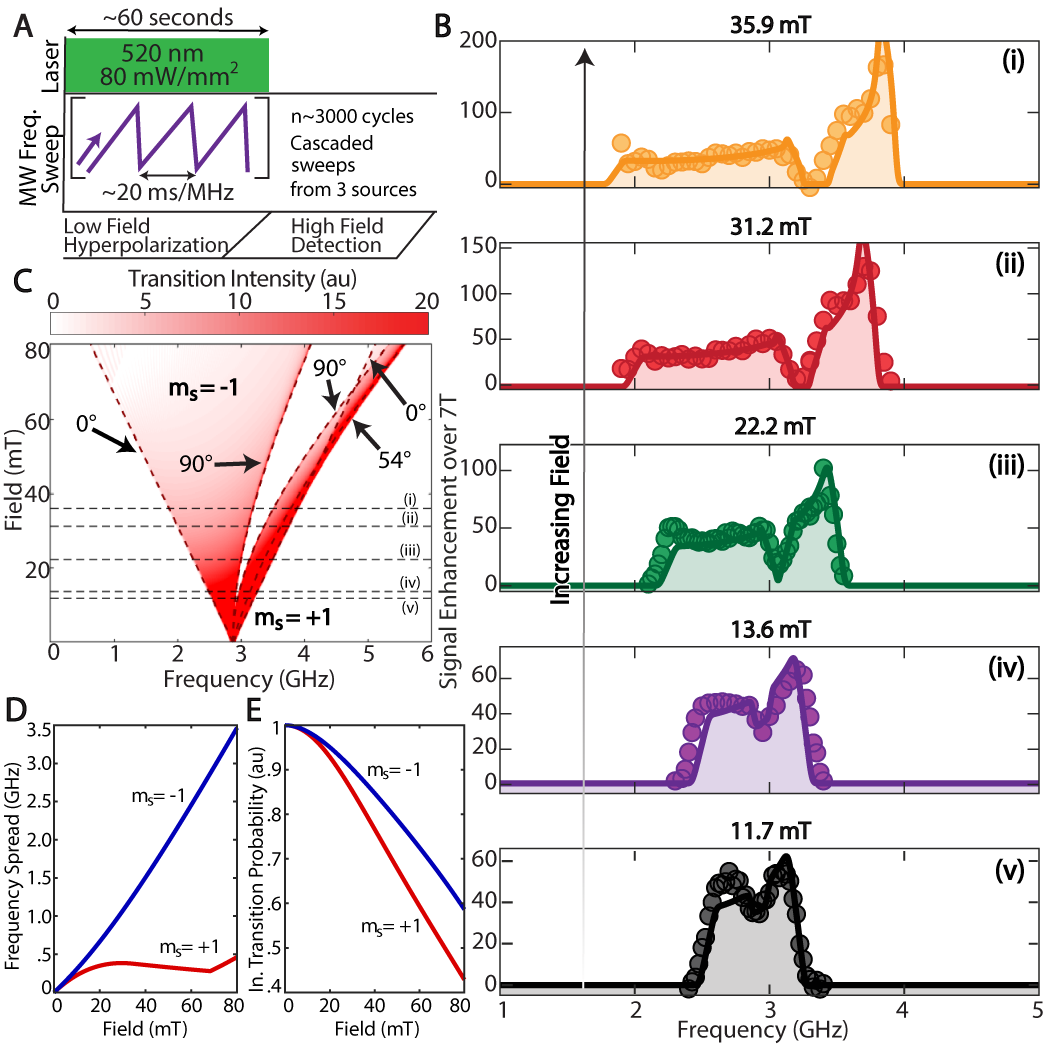}}
  \caption{\T{Fully orientation independent hyperpolarization.} (A) \I{DNP pulse sequence} at low-field involving simultaneous laser and chirped MW irradiation over the NV center ESR spectrum. DNP enhancements are quantified by NMR detection at 7T. (B) \I{Electronic spectra mapped via $\Cs$ DNP} on 200$\mu$m particles (\zfr{prototype}C). Hyperpolarized $\Cs$ NMR is performed on narrow 100MHz frequency windows (points) to map the underlying NV ESR spectrum at various fields \I{(i-v)}. Solid lines are fits to calculated spectrum (dashed linecuts in C) and show remarkable agreement. Panels illustrate that \I{all} orientations of the particles contribute constructively to hyperpolarization, and all with the same sign. (C)  \I{Simulated NV electronic spectra}, shown as a function of field, due to varying orientations ($\xt$) of N-V axes in crystallites in the diamond powder with respect to the polarizing field $\Bpol$. Shading is proportional to electronic transition intensities $P(\xt,\Bp)$. Exact orientations that contribute to the extrema of the patterns in $m_s=\pm 1$ manifolds are indicated by arrows and the red dashed lines.  (D) \I{Frequency spread} in either electronic manifold, showing an approximately linear fan out of transitions in the $m_s=$-1 branch, as opposed to being constant ($\app$400MHz)  in the $m_s=$+1 branch at moderate fields. (E) \I{Integrated transition intensity} ${\cal T}(\Bp)$ averaged over all orientations of the powder, indicating that the net transition intensity falls with increasing field.}
 \zfl{theory}	
\end{figure}

\T{\I{Fully orientation independent hyperpolarization:}} -- In Ref.~\cite{Ajoy17}, we had observed that every part of the NV electronic spectrum contributes to DNP, and that the sign of the resulting $\Cs$ polarization depends on the direction of the sweep. While this was a strong indication that DNP was excited for all orientations, it was not immediately evident that polarization builds up \I{equally} efficiently in \I{every} orientation of random crystallites in the diamond powder. In this paper, we perform new experiments that quantitatively answer this question in the affirmative, making our polarization mechanism perhaps the first reported optical-DNP process that is completely orientation independent while also being constructive over the full electronic bandwidth. It also stands in contrast to other proposals for nanodiamond DNP~\cite{chen15b}, where only spins in a narrow cone are polarized. These results also point to simple means to \I{optimize} hyperpolarization enhancements at any given polarizing field $\Bp$.

We perform experiments sweeping MWs in narrow 100MHz windows, using the obtained $\Cs$ DNP enhancements to report on the underlying NV ESR spectra~\cite{Ajoy17}. We employ a Helmholtz coil (see \zfr{field}D) within the polarizer device to generate varying polarization fields in a background fringe field of a 7T NMR spectrometer. Embedded chip-scale Hall sensors measure the field along three axes, and the vector fields are reported in \zfr{theory}B. As expected the spectra become wider at higher fields. The key new result in this paper, however, is contained in the solid lines that provide remarkably good fits to the experimental data. These fits involve a single free parameter (overall amplitude), and are derived from a simple model of the NV center electronic spectrum alone, without having to include hyperfine couplings to the $\Cs$ nuclei. The experiments reveal therefore that the $\Cs$ hyperpolarization fully \I{follows} the underlying NV electron density of states, pointing to \I{``complete''} orientation independence. They also reflect that it is the weakly coupled $\Cs$ nuclei that are predominantly polarized.

To be more specific, \zfr{theory}C shows simulated NV ESR spectra at various fields, with the shading color being proportional to the transition intensity under applied MW excitation. The solid lines in \zfr{theory}B are line-cuts in this graph (dashed lines). We first start with the spin-1 NV center Hamiltonian $\mH(\xt) = \xD S_z^2 + \xg_e\Bp(S_x\sin\xt + S_z\cos\xt)$, where $\xD$=2.87GHz is the zero-field splitting, $\xg_e$=28MHz/mT is the electron gyromagnetic ratio, and $S_j$ are Pauli matrices. We calculate the eigenvectors such that $\mH\ket{v_k}=E_k\ket{v_k}$, and obtain the transition intensities $P(\xt,\Bp)\sim\sum_{k<\ell}\sum_{m}|\sand{v_k}{S_m}{v_{\ell}}|^2[\sand{v_k}{\xr}{v_k} - \sand{v_{\ell}}{\xr}{v_{\ell}}]$, where the first factor quantifies the transition probabilities in the randomly oriented powder where $m\in\{x,y,z\}$. The second factor describes the population difference between the eigenstates, with $\xr=\id-S_z^2/3$. The predicted spectra are then calculated by assuming a Gaussian spectral width $\app$28MHz (corresponding to a field inhomogeneity $\app$1mT) for each transition, averaging the effective $P(\xt)$ over 300 random orientations in the powder, and then convoluting the result by the sweep window. The extremities of the spectra at low fields are easy to identify, at frequencies $\xD\mp\xg_e\Bp$, as originating from the crystallites aligned with $\Bp\: (\xt$=0$^{\circ}$) in the $m_s=\mp 1$ manifolds. Similarly the perpendicular ($\xt$=90$^{\circ}$) orientations occur more centrally in the powder pattern, at frequencies $\fr{1}{2}[\xD + \sqrt{\xD^2 + (2\xg_e\Bp)^2}]$ and $[\sqrt{\xD^2 + (2\xg_e\Bp)^2}]$ respectively. The region in-between the two manifolds has no electron density of states and consequently produces no DNP enhancements, resulting in the apparent ``holes'' around 3GHz in \zfr{theory}B.

This simple model now immediately opens the door to ways to optimize the DNP enhancement. Firstly, knowledge of the \I{vector} polarizing field, for instance through Hall probes embedded in the device near the sample, can point to the exact frequency band $\mB$ for the MWs to sweep over. This ensures that the applied microwaves are sweeping over electrons at \I{every} time instant during the full polarization period (constrained by nuclear $T_1$). Moreover, as \zfr{theory}D demonstrates, the frequency spread in the $m_s$=-1 manifold grows approximately linearly with field, while in the $m_s$=+1 manifold it saturates after an initial quadratic rise. Since there is a \I{relative} reduction of electron density of states per unit frequency bandwidth in the $m_s$=-1 manifold as opposed to the $m_s$=+1 manifold, higher DNP enhancements are obtained by MW sweeps over the $m_s$=+1 branch. This is also evident in the experiments in \zfr{theory}B. At 36mT, for instance, a single 100MHz sweep window in the $m_s$=+1 manifold can provide DNP enhancements approaching 200 over 7T (\zfr{theory}B).

Finally the excited DNP is as a result very \I{robust} to generate. \I{(i)} Any part of the frequency band can be swept over to produce hyperpolarization, in contrast to conventional DNP (solid/cross-effects), where misplaced frequency windows can lead to destructive polarization generation between various spin packets. \I{(ii)} Given that the ESR spectrum is orientationally broadened to start with, field inhomogeneities do not significantly alter the DNP enhancements. We estimate a inhomogeneity of 2mT over the sample volume (see Supplemental Information \cite{SOM}) in our device. \I{(iii)} Moreover, since the NV electrons are quantized along the randomly oriented axes in the powder, $\Bpol$ can be applied in \I{any} direction. It can hence be generated by a \I{vector} combination of a single-axis Helmholtz coil and the magnet fringe field. These favorable settings enable the simple installation of the device in the vicinity of detection magnets -- one simply ``dials-up'' the current in the Helmholtz coil such that the net vector field seen by the sample is the optimal value $|\Bpol|\app$38mT.

\begin{figure}[t]
  \centering
	{\includegraphics[width=0.48\textwidth]{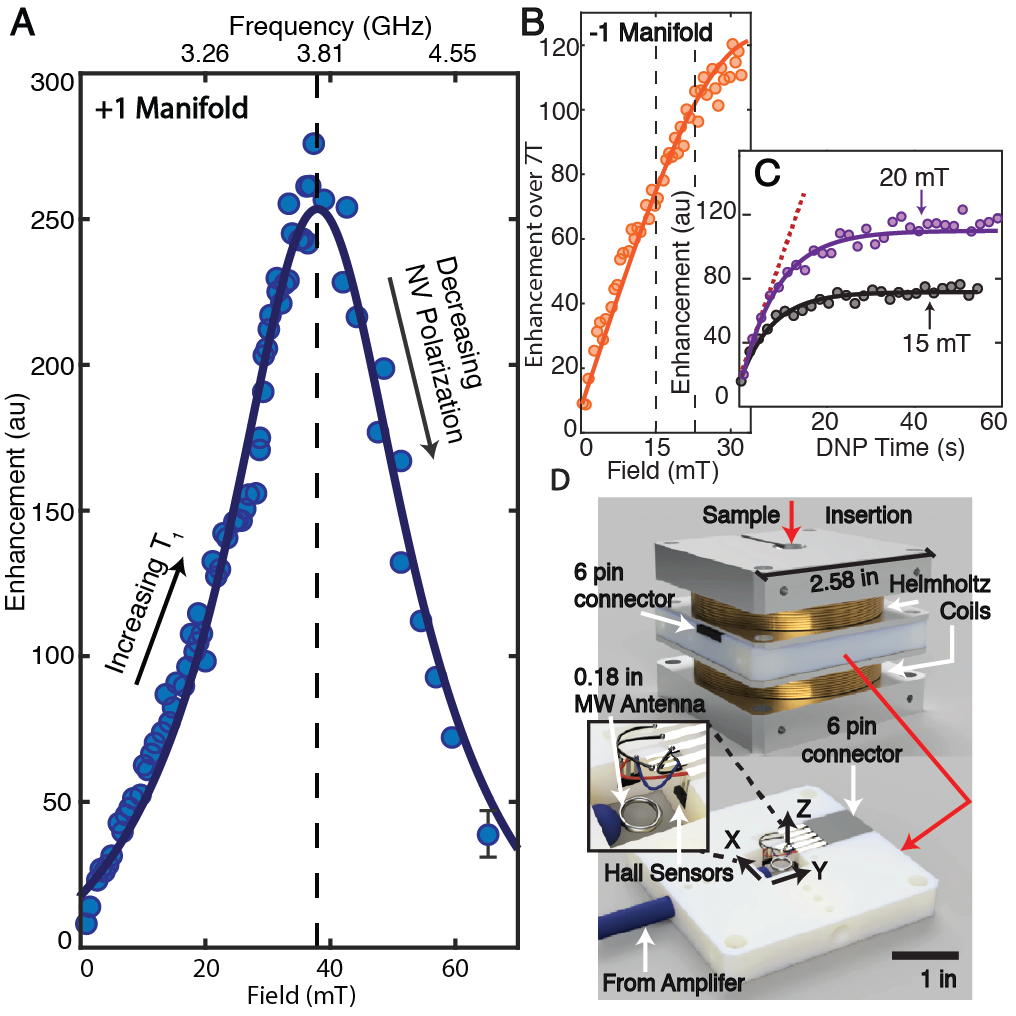}}
        \caption{\T{Field dependent DNP profile.} (A) \I{Field dependence} of maximum hyperpolarization enhancements in 1-70mT range for the $m_s$=+1 NV manifold. Experiments are performed at optimal MW sweep rates at each field (see \I{Methods}). We observe a steep decrease in DNP efficiency at ultralow fields $<$20mT and at fields $>$50mT, and a sharp optimum at $\app$38mT.  We ascribe this to be due an interplay between low $\Cs$ lifetimes at low field, and inefficient optical polarization in a powder at high fields. Center of MW sweep bands are shown on top axis.  Solid line is guide to the eye. Error bar $\app$11\% is shown representatively on last point. (B) Similar experiments performed on $m_s$=-1 manifold.  (C) Comparison of polarization buildup curves at 15mT and 20mT, showing longer saturation times for the latter, a reflection of longer nuclear relaxation times. Red dashed line shows identical rate of polarization buildup at both fields. (D) \I{Helmholtz coil} placed over the MW loop antenna onboard nanodiamond hyperpolarizer device (see also \zfr{model}) can be used to apply the polarizing field in a variety of environments of NMR detection magnets. \I{Inset:} Three Hall probes measure the \I{vector} magnetic field in close proximity to the MW antenna and sample.}
 \zfl{field}	
\end{figure}

\T{\I{Low-field only hyperpolarization:}} -- We now turn our attention to the field dependence of  the DNP enhancements. Determining an \I{apriori} analytical model from microscopics of the ``DNP ratchet''~\cite{Ajoy17, Zangara18} is extremely challenging since there are several factors at play simultaneously: different orientations in the powder, a continuum of hyperfine couplings within the diamond lattice, as well as always operational nuclear spin diffusion and relaxation effects. However one could generally state that the DNP mechanism is operational at low fields, since the critical hierarchy that we rely on, $\xo_L<|A|$ flips at higher fields. Indeed in this regime, the mechanism will transition to the more conventional Integrated Solid Effect (ISE)~\cite{Henstra14}, with several contrasting features.

Before going forward, we do emphasize however, that these low-field regimes ( $\lesssim$70mT) have been traditionally \I{inaccessible} to DNP since typically electron polarization is also generated through a Boltzmann distribution at cryogenic conditions at high fields. Indeed, quantum materials such as the NV center provide a new paradigm on account of the fact that the electronic spins can be polarized optically \I{independent} of temperature and even at \I{zero} magnetic field. A hint of the possible field dependence (neglecting nuclear relaxation) is elucidated in the simulations of \zfr{theory}E. We plot here the integrated transition intensity of the electron spectra as a function of magnetic field, ${\cal T}(\Bp) = \int_0^{\pi/2}P(\xt,\Bp)$. This suggests that the available NV density of states for hyperpolarization reduces with increasing field. 

In this paper we provide an experimental solution, reporting in \zfr{field} the measured field dependence of the obtained DNP enhancements under optimal conditions. 
We use a combination of Helmholtz coil fields and background fringe fields over a wide range (1-70mT) to systematically map the field dependence under sweeps of the $m_s$=+1 (\zfr{field}A) and  $m_s$=-1 manifolds (\zfr{field}B) respectively. In actuality, due to amplifier constraints, the data in \zfr{field}A is obtained in two separate data sets (1-40mT) and (30-70mT) and pieced together by normalizing overlapping points. We estimate from this an $\app$11\% estimate through this (marked in the last point of \zfr{field}A). We see a remarkably sharp field dependence, becoming optimal around $\Bp$=38mT$\pm$4mT, and falling steeply on either side of this value. 

While a quantitative model is still beyond the scope of this manuscript, we ascribe this behavior to be arising from a competition between two factors with increasing $\Bp$: \I{(i)} a dominant rise in $\Cs$ nuclear $T_1$ lifetimes and \I{(ii)} a fall in the NV center polarization~\cite{Goldman15} and integrated electron transition probability ${\cal T}(\Bp)$ in randomly oriented diamond powder. The nuclear $T_1$ is set for the most part (see \zfr{T1}) by interactions with the spin bath of paramagnetic electron defects (primarily P1 centers) in the diamond lattice, which present to the $\Cs$ nuclei a spin-flipping noise spectral density centered at zero frequency and width given by approximately the inter-electron dipolar coupling. Increasing field allows the $\Cs$ nuclei to sample less of this noise, leading to an increase in $T_1$ and the ability to buildup polarization for longer times before saturation. This is demonstrated in \zfr{field}C (\I{inset}) where we report DNP buildup curves at 15mT and 20mT, and where the polarization curves saturate at longer times in the latter case. Simultaneously however, there is a reduced overall NV transition probability ${\cal T}(\Bp)$ at higher fields (\zfr{theory}E), and also (ignored by \zfr{theory}E) a reduction in number of $\Cs$ nuclei directly participating in the DNP process (satisfying the hierarchy $\xo_L<A$). Due to these factors we expect a decrease in DNP efficiency at high fields. 

Due to these competing factors, we expect that the exact optimal field is sample dependent. Overall however, the low polarizing fields $\Bp\app$40mT, and the relatively benign range around this field $\xD\Bp\app$10mT, mean that they are simple to generate through permanent magnet or coils. This feature is \I{key} to miniaturization of the hyperpolarizer. Moreover the MW bandwidth in the optimal $m_s=$+1 manifold to sweep over is relatively narrow $\mB\app$0.35GHz around 3.81GHz. We note that, in contrast, at high fields $\Bp\gg$100mT, where the mechanism transitions to the standard ISE, one has to contend with far reduced electron density of states and the $2\xD\sim$5.9GHz wide electron spectral width which is technologically challenging to sweep over.

\begin{figure}[t]
  \centering
	{\includegraphics[width=0.48\textwidth]{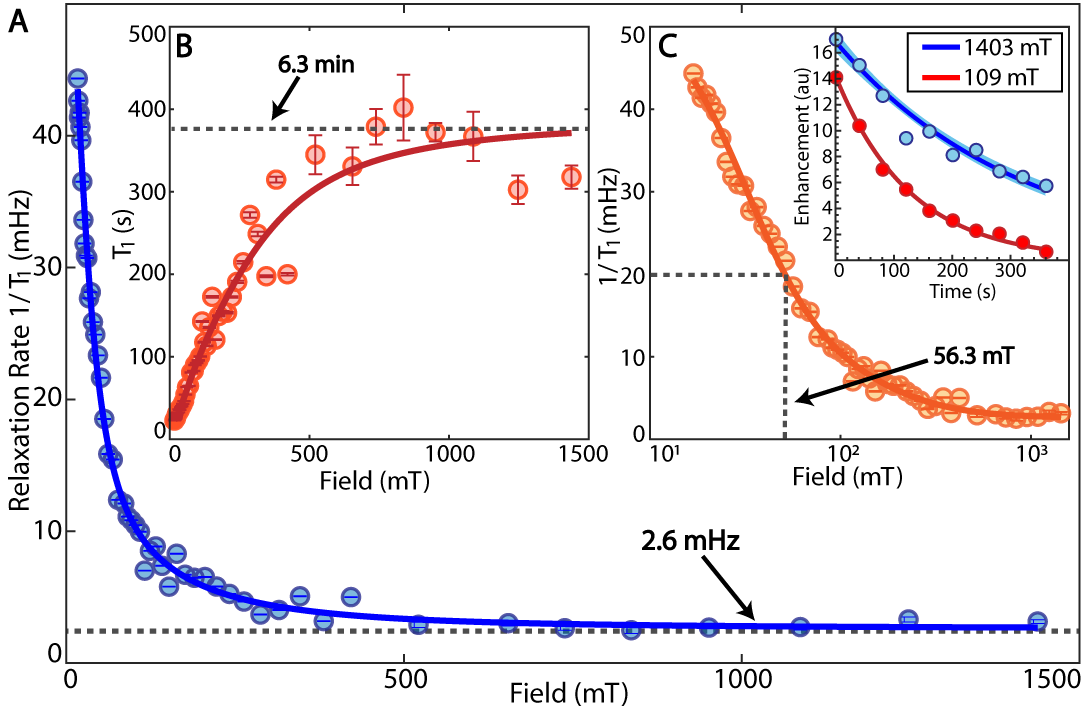}}
  \caption{\T{Lifetimes of hyperpolarized $\Cs$ nuclei}. (A) \I{Field dependence} of nuclear relaxation rate $R_1=1/T_1$ for hyperpolarized 200$\mu$m particles in \zfr{prototype}C. There is a steep rise in relaxation rate at low-fields below $\lesssim$100mT mediated by interactions with paramagnetic impurities in the lattice. At higher fields the rate is approximately constant $\app$2.6mHz, allowing efficient retention of hyperpolarization for minute-long periods. Solid line is a Lorentzian fit. Error bars are estimated from monoexponential fits (see \I{Methods}~\cite{SOM}). (B) Measured $T_1$ values showing lifetimes $>$6min at modest fields. (C) \I{Knee field} at which the rapid increase of nuclear lifetimes occurs can be quantified as the width in a logarithmic scale, here $\app$57mT. \I{
Inset:} Signal decays at exemplary low and high fields. Shaded are 95\% confidence intervals.}
	\zfl{T1}
\end{figure}

\T{\I{Long time hyperpolarization retention:}} --  While the DNP mechanism is optimum at low fields, we also find that modestly low fields are sufficient to retain this polarization for long periods, with typical cases approaching ten minutes. We demonstrate this in \zfr{T1} by performing a full wide range ($\Br$=10mT-7T) field dependent mapping of the $T_1$ relaxation of $\Cs$ nuclei in typical diamond microparticles (\zfr{prototype}C). This is achieved by retrofitting a field cycling instrument constructed over a 7T magnet~\cite{Ajoyinstrument18} with our optical hyperpolarizer device. One would naively expect a increase in relaxation rate that falls down as $R_{1}\propto 1/\Br$, due to a suppression in electron-nuclear overlaps due to the widening energy gap between the two reservoirs. We find, however, a dramatic \I{step-like} dependence on field (see \zfr{T1}A). There is a strong increase in $T_1$ beyond a particular \I{knee} field $\app$57mT where lifetimes approach 6.3min and a steep fall below this field (see also \zfr{T1}B). Knee fields can be most easily quantified in a logarithmic field plot of the relaxation plots (\zfr{T1}C).  We have observed that the dependence in \zfr{T1} is typical of diamond particle samples employed for hyperpolarization. We explain this behavior to be arising from the interaction of the $\Cs$ nuclei with paramagnetic impurities, the knee field value being a dominant function of the P1 center concentration. As we shall present in detail in a a forthcoming manuscript with more detailed experiments and analytical models, these trends are general across all types of samples employed for hyperpolarization.

That said the step-like dependence in \zfr{T1}A immediately opens the door to enormous simplification of hyperpolarizer operation and deployment:  by rapidly switching the field to $\gtrsim$100mT after optical pumping by means of an electromagnet, one could retain the polarization for minute-long periods. Moreover the behavior in \zfr{T1}A ensures that the hyperpolarization loss during sample shuttling can be exceedingly small ($<$1\% in our experiments), since it is only the traversal time through ultra-low field regions that predominantly contribute to deleterious loss.

\begin{figure}[t]
  \centering
	{\includegraphics[width=0.48\textwidth]{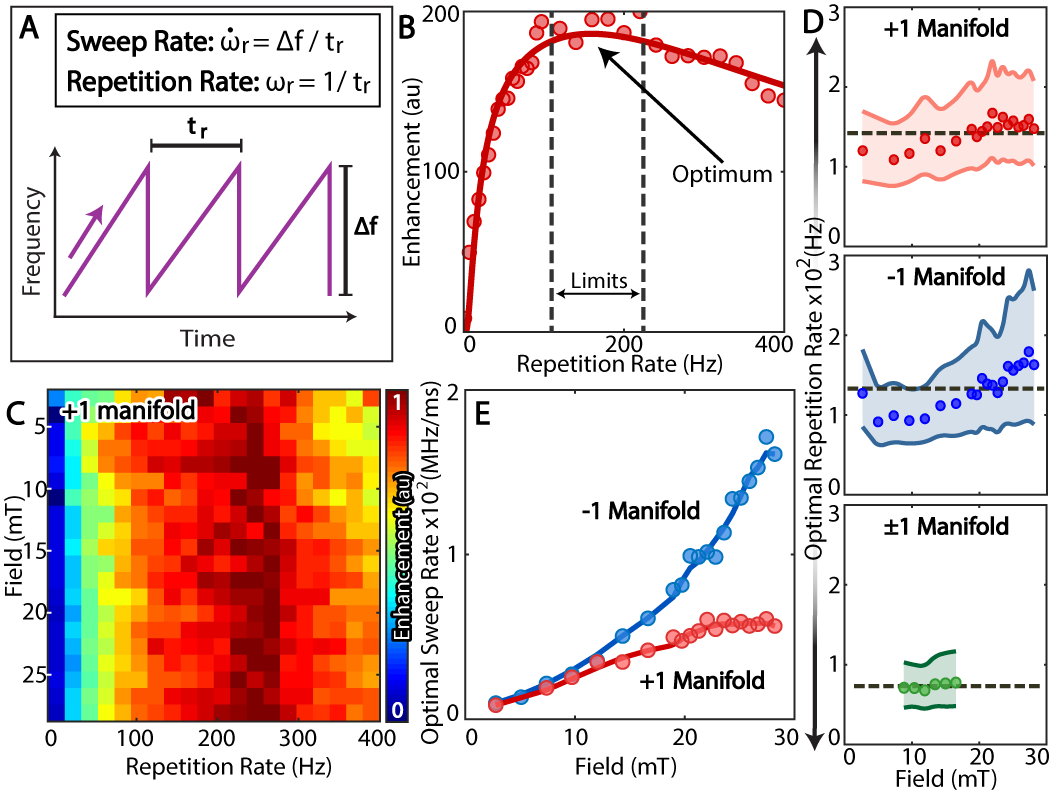}}
  \caption{\T{MW sweep rates dependence} of hyperpolarization enhancements. (A) \I{Definitions.} We distinguish between MW repetition rates $\xo_r$ and sweep rates $\dxo_r$.  (B) Exemplary variation with MW repetition rate at 28.8mT employing three cascaded MW sources sweeping a bandwidth $\mB$=384MHz. Solid line denotes a fit to microscopic model~\cite{Ajoy17} that qualitatively captures the observed behavior, with dashed lines denoting 5\% confidence interval. (C) \I{Full dataset} as function of field for MW sweeps over the $m_s$=+1 manifold. Color reflects obtained 7T DNP enhancements. (D) \I{Optimal MW repetition rates} $\xo_r$  measured for sweeps over the $m_s=\pm$ 1 manifolds separately or together. Optimal value is obtained from fits. Remarkably, data reveals relative field independence of observed optimal rates, with $\xo_r^{(+1)}\app$147Hz, $\xo_r^{(-1)}\app$133Hz, and  $\xo_r^{(\pm1)}\app$73Hz (dashed lines). Shaded area represents 5\% confidence interval around the optimal rates. Data suggests optimal repetition rates are dominated by the need to sweep each NV once during every optical repolarization cycle (E) \I{Optimal sweep rates} $\dxo_r$ field dependence plotted for sweeps over the $m_s=\pm$1 manifolds individually. Data follows the spread of the underlying electronic spectra in both manifolds (see \zfr{theory}D). }
 \zfl{sweep}	
\end{figure}

\T{\I{Factors simplifying MW sweeps:}} -- We now focus our attention on the factors affecting MW sweeps through the optimal electron bandwidths. For simplicity (see \zfr{sweep}A), we distinguish between MW \I{sweep rates} $\dxo_r=\mB/t_r$,  where $t_r$ is the time per sweep, and \I{repetition rates} $\xo_r=1/t_r$. which are instead the rate of frequency chirps through the sweep band $\mB$. \zfr{sweep}B shows a typical dependence on the repetition rate, demonstrating a loss of polarization efficiency at slow and fast rates.

Microscopic predictions of the optimal MW sweep rates, and their dependence on polarizing magnetic field $\Bp$, are once again extremely challenging. The Landau-Zener energy gaps and consequently the conditions for adiabatic travels of the level anti-crossings are functions of the applied Rabi frequency $\xO_e$, NV center orientation, and importantly the hyperfine couplings~\cite{Ajoy17}; and given the continuum of couplings and orientations in our sample the position of the optimum is difficult to analytically compute. However, as \zfr{sweep}B demonstrates, knowledge of the sweep rates is critical to obtaining optimized hyperpolarization enhancements. 

In this paper, we experimentally address this question, by determining dependence on MW sweep rates for various polarizing fields. \zfr{sweep}C shows these results, concentrating first on the $m_s$=+1 manifold, and employing a frequency comb constructed out of three cascaded MW sweepers. For each polarizing field, we sweep over the full NV ESR band $\mB$ given by \zfr{theory}C. Surprisingly, we find that the MW \I{repetition} rates $\xo_r^{(+1)}$ are (to a good approximation) \I{independent} of magnetic field. To demonstrate this more clearly, in \zfr{sweep}D, we extract the optimal repetition rates for fields $\Bp$=1-30mT, considering sweeps over the $m_s$= +1 and -1 manifolds separately, as well as over both manifolds together. These optimal values are obtained from fits of the observed dependence (eg. solid line in \zfr{sweep}B) to the expected behavior from microscopics of the Landau-Zener process~\cite{Ajoy18}, $\varepsilon=A\exp(-\xL^2/\xo_r)(1-\exp(-\xO^2/\xo_r))$. The lines in \zfr{sweep}D indicate 95\% confidence intervals, showing additionally that the repetition rates exhibit a relatively benign dependence, with width $\xD\xo_r\sim$50Hz. In contrast, by plotting the MW sweep rates $\dxo_r$ in \zfr{sweep}E, we find that they increase with field, and closely follow the underlying spread in the electronic density of states (see \zfr{theory}D).

It is confounding that the complex system microscopics conspire to produce relative field independence. Strong hints to the origin of this behavior is provided by the results in \zfr{sweep}D, where we observe that the optimum repetition rates for sweeps over both $m_s=\pm$1 manifolds simultaneously $\xo_r^{(\pm 1)}\app$73Hz$\pm$29Hz, is approximately \I{half} that of sweeps over the individual manifolds, for instance, $\xo_r^{(+1)}\app$147Hz$\pm$53Hz. Indeed, several factors contribute to determining optimal MW rates. There is the need to maximize: \I{(i)} the polarization transfer efficiency \I{per} sweep \I{(ii)} the \I{total} number of sweeps in a period bounded by nuclear $T_1$, and \I{(iii)} and the NV electron polarization at every sweep event. Experiments in \zfr{sweep}B suggest that the last factor is the most critical, pointing to relative field independence. Indeed, at the laser powers we employ the NV repolarization rate $\trepol\sim T_{1e}\app$1ms~\cite{Jarmola12}, and sweeping MWs at a rate $\xo_r\app 1/(NT_{1e})$, ($N$=3 being the number of sweepers) ensures the largest NV polarization is available to transfer to the $\Cs$ nuclei per sweep.

This surprising aspect once again simplifies hyperpolarizer miniaturization. Sweep times (5-10ms) are relatively slow, and can easily be generated by using microcontrollers to provide voltage ramps that when interfaced with the chip-scale VCOs provide the frequency chirped MWs (see Supplemental Information~\cite{SOM} for a miniaturized custom-built frequency-chirp circuit used in our hyperpolarizer). Indeed miniaturization is the key behind the ability to cascade multiple frequency sources in a frequency comb in order to provide multiplicative enhancement gains~\cite{Ajoy18}.

\begin{figure*}[t]
  \centering
  {\includegraphics[width=0.75\textwidth]{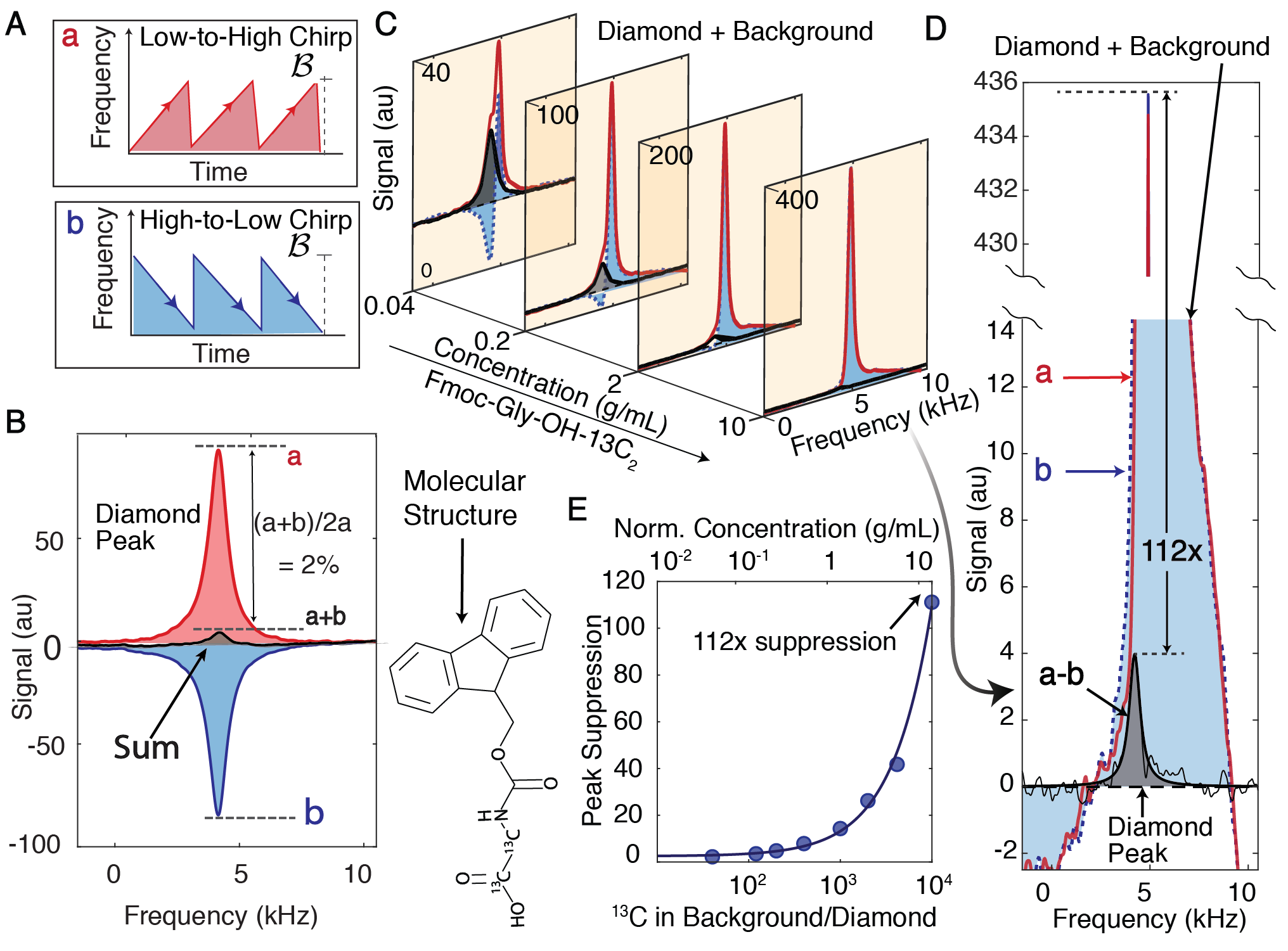}}
  \caption{\T{``\I{Background suppressed}'' $\Cs$ hyperpolarization} employing polarization sign reversals. (A) \I{Schematic of polarization sign control.} Sign of $\Cs$ hyperpolarization, aligned or anti-aligned to $\Bp$, depends only on the direction of the MW sweep. (B) \I{Sign reversal fidelity} $\cal F$ is evaluated by MW sweeps at $\Bp=$22mT. Results demonstrate that hyperpolarization sign can be reversed on-demand to better than (1-${\cal F}$) = 2\%.  (B) \I{Background suppression} by exploiting successive sign-reversals of $\Cs$ hyperpolarization. Diamond particles in \zfr{prototype}C are immersed in increasing concentration (\I{panels}) of Fmoc-Gly-OH-$\Cs_2$ (\I{``background''}) that overlaps the diamond spectrum. Red (blue shaded) line is the obtained spectrum under low-to-high (high-to-low) MW sweeps, each averaged 20 times. Subtracting the results allows one to extract the diamond spectrum (black shaded) although initially indiscernible. Concentrations of the background compound are displayed normalized to 50 diamond particles (see \I{Methods}). (D) \I{Zoomed} signal when employing highest relative concentration (final panel in C). Diamond signal (black) is recovered with high fidelity although initially enveloped by a 112 times stronger background signal, which is now completely suppressed. (E) \I{Scaling of background suppression} with compound concentration (upper axis). In final panel of C, suppression exceeds two orders of magnitude, corresponding to the detection of diamond $\Cs$ nuclei immersed in $\sim$10$^4$ more $\Cs$ nuclei in the background (lower axis). }
 \zfl{imaging}
\end{figure*}

\begin{figure}[t]
  \centering
	{\includegraphics[width=0.49\textwidth]{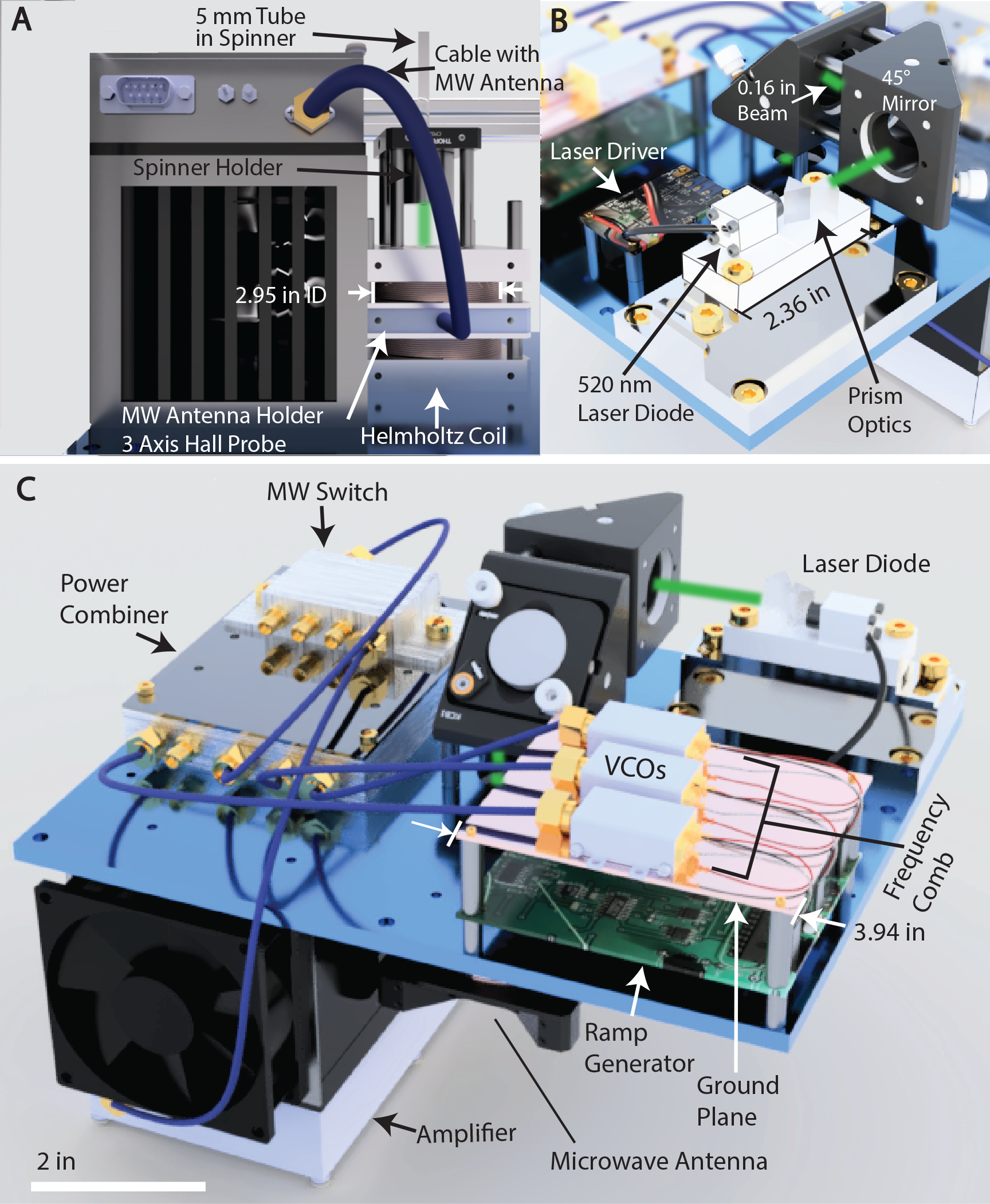}}
  \caption{\T{Nanodiamond hyperpolarizer schematic} (to scale) showing component parts assembled on a single 10in$\zt$10in aluminum plate (blue). (A) \I{Sample attachment}. Diamond particles to be hyperpolarized are placed at the confluence of the laser and MW excitation in a weak polarizing field generated by the Helmholtz coil pair. Amplified MWs are delivered by means of a stubbed loop antenna (see \zfr{field}D). (B) \I{Optical excitation} is provided by a small (2.36in) 520nm 1W laser diode placed on the plate underside. Minimal optics direct the beam to the sample. Alternatively, multi-mode optical fibers can deliver the light from one or more laser diodes mounted in the device. (C) \I{MW generation and excitation} circuitry (also on plate underside) involves miniature VCO sources interfaced with a custom-built voltage ramp generator to produce the MW frequency sweeps. Chirped MWs are subsequently power combined, amplified, and finally delivered to the MW antenna. The ultracompact, modular, design allows easy customization and enhances device portability.}
 \zfl{model}	
\end{figure}

\T{\I{``Background suppressible'' hyperpolarization:}} --  Finally another striking feature of the DNP mechanism allows additional SNR gains in the detection of hyperpolarized signals, especially in the presence  of large background signals. We refer to ``\I{background}'' as those signals that are not arising directly or indirectly from the optically hyperpolarized diamonds. We exploit the observation that the sign of the $\Cs$ polarization depends only upon the direction of the MW sweep~\cite{Ajoy17}. Polarization is aligned (anti-aligned) to $\Bpol$ for low-to-high (high-to-low) frequency sweeps over the NV ESR spectrum (\zfr{imaging}A). In this paper, we quantify the \I{inversion fidelity} ${\cal F}$ of the sign reversals. Remarkably, we find that when all NV orientations are swept over, the sign reversal is extremely robust, the amplitude of the $\Cs$ polarization inverts to within $(1-\cal{F})$ = 2\% of its original value (see \zfr{imaging}B). In reality, this is an underestimate of the inversion fidelity since it also includes repolarization effects during sample shuttling.

This powerful ability to invert hyperpolarization signals on-demand and at high fidelity, with no change of hardware infrastructure, opens the door to \I{background suppression} of the DNP signals. Performing successive experiments with alternate MW sweep direction and subtracting the result, one can suppress any background signals and exclusively recover the DNP signal although it maybe initially impossible to discern. Applications for this idea are particularly compelling if the polarization from the $\Cs$ spins in the diamond powder can be transferred to external liquids, for example $\Hs$ spins in water, through cross polarization~\cite{Ji17} or the Overhauser effect~\cite{Navon96}. Such hyperpolarized water, with a polarization that is sign invertible at will,  could then be used as a bright-field MRI contrast agent in imaging applications, completely suppressing $\Hs$ signals from thermally polarized water in the body, and significantly boosting image SNR.

In this paper, we demonstrate a restricted proof-of-concept experiment along these lines. We perform DNP on diamond microparticles embedded in a large volume of $\Cs$ labeled Fmoc-Gly-OH-$\Cs_2$, a compound~\cite{Harris00} with a chemical shift that completely overlaps with the diamond signal at 7T. In \zfr{imaging}C, we increase the concentration of this ``background'' signal, so that the hyperpolarized diamond signal is completely enveloped by it and typically impossible to discern. In fact, in the final panel of \zfr{imaging}C, the background signal is over hundred times larger than the diamond peak. However background suppression using polarization sign inversions works remarkably well (see \zfr{imaging}C), and we can extract the diamond peak from the 112 times larger background (zoomed in \zfr{imaging}D) with high fidelity.

An instructive representation of SNR gains through a \I{combination} of hyperpolarization and background suppression is in the lower axis of \zfr{imaging}E. We are able to discern the diamond peak with high SNR, although every $\Cs$ nucleus in diamond is immersed in a background of more than $10^4$ more $\Cs$ nuclei. This large factor is a multiplicative effect of DNP that renders every $\Cs$ in the diamond roughly 100 times ``brighter'' than 7T thermally polarized nuclei, and polarization sign-reversals that suppress the $\Cs$ signatures in the background by over 100 times. The high fidelity recovery of the hyperpolarized signal in \zfr{imaging}D indicates that the trend in \zfr{imaging}E can be continued further; indeed we in the ultimate limit one could suppress background signals over few more orders of magnitude.

\T{\I{Conclusions:}} -- In conclusion,  by uncovering new physics underlying the DNP mechanism first introduced in Ref.~\cite{Ajoy17}, we have constructed the first compact solid-state room-temperature ``\I{optical nanodiamond hyperpolarizer}''. Indeed the ability to construct such a device is \I{itself} surprising. It arises from a unique confluence of factors underlying the DNP mechanism requiring 
low fields, low laser and MW powers, and being robust to field inhomogeneity, optical excitation modes, and MW inhomogeneity. Hyperpolarization is on-demand sign invertible with high-fidelity, can be excited with very modest resources, and retained for long periods approaching tens of minutes. We employed the device to obtain best reported values of $\Cs$ hyperpolarization in diamond micro- and nanoparticles through optical means. We have also highlighted engineering aspects that, leveraging the physics, make our nanodiamond hyperpolarizer device easy to build and operate, and with a small footprint that can retrofit any existing magnet system.

Our work opens the door to many intriguing new future directions. \I{First}, variants of the hyperpolarizer device could enable efficient polarization transfer relayed from the long-lived surface $\Cs$ nuclei to external liquids. Diamond nanoparticles could be made to dress walls of a narrow capillary through which liquid is pushed through to be hyperpolarized. This hyperpolarization so generated optically is replensible, and could also be employed with NV sensors for the optical detection of magnetic resonance~\cite{Ajoy2015,Aslam17,Glenn18}. This will open new possibilities for miniature NMR spectrometers for chemical analysis.  \I{Second}, from a technological standpoint, the device can be easily miniaturized further. A more efficient MW delivery scheme will allow the use of a lower power amplifier that currently occupies the largest footprint. All other MW components can be replaced by chip-scale ones, and a palm-top sized hyperpolarizer can easily be envisioned. Integrating low-field inductive NMR readout~\cite{Lee08} into the device will enable an in-situ measurement of the $\Cs$ hyperpolarized signals without the need for sample shuttling.  \I{Finally}, it should be possible to produce similar low field hyperpolarization in other wide bandgap semiconductor materials.  Pioneering recent work~
\cite{Falk15,Nagy18} has  demonstrated that the V1 defect center in Silicon Carbide can be hyperpolarized through infrared light. Employing these spins for DNP will allow \I{in-vivo hyperpolarization} of $\Sis$ nuclei and background-free imaging of nanoparticles targeting disease locations.

\T{\I{Acknowledgments:}} -- We gratefully thank D. Budker, B. Bl\"{u}mich, S. Conolly,  A. Gali, F. Jelezko, M. Lustig, C. Ramanathan, D. Sakellariou, O. Shenderova and J. Wratchrup for insightful conversations. We acknowledge technical contributions from X. Cai, S. Le, G. Li, A. Lin, X. Lv, T. McNelley, P. Raghavan, I. Yu,  and R. Zhao. Correspondance and request for materials should be addressed to A.A. (\I{ashokaj@berkeley.edu}). 

\T{\I{Materials and Methods:}}  

\I{Hyperpolarizer construction:} -- The hyperpolarizer (\zfr{prototype}A) is a stand-alone device fully operational at room-temperature, composed from nonmagnetic solid-state components and requiring zero user maintenance. We choose a modular design that allows a compact and rapid assembly of the various components (see \zfr{model}). The most striking feature is its small footprint (12$\zt$10$\zt$10in.), and light weight ($<$10lb), making it ultraportable, and compatible with any NMR spectrometer. We believe this to be the smallest reported hyperpolarizer across all platforms, a testament to the technological ease of optical DNP at low fields.

The aluminum chassis supports three distinct modular blocks: optics, MW sweep generation, and a sample holder that contains the diamond particles to be hyperpolarized (\zfr{model}). The densely packed and double-sided design supports easily customizable modalities for sample placement and removal. For instance, the device can contain a hollow bore to allow shuttling of the sample into a high field NMR magnet. We employ a miniature 1W 520nm diode laser (Lasertack PD-01289) in a feedback loop with an integrated thermoelectric cooler for adequate thermal control (TE Inc. TE-63-1.0-1.3). Very few optical components are required (see \zfr{model}B): an aspheric lens and a set of anamorphic prisms collimate the beam to a circular 4mm diameter. Two mirrors redirect the beam towards the sample, typically irradiating it from below. In addition, to polarize larger sample masses, one could interface multiple fiber coupled (Thorlabs M35L01) laser diodes to excite the sample also from the sides.

Microwaves are generated by miniature voltage controlled oscillator (VCO) sources (Minicircuits ZX95-3800A+, 1.9-3.7GHz, output power $p$ = 3.1dBm). Frequency sweeps are produced by controlling the VCO frequency by a homebuilt quad-channel voltage ramp generator controlled by a PIC microprocessor (PIC30F2020). \zfr{model}C shows the connectorized VCOs mounted on a copper sheet that serves as a good ground plane. Given the relatively slow MW sweeps required, $\xo_r\app$164Hz, translating to sweep times of 6ms for the typical sweep bandwidths $\mB$=100MHz-1GHz, the 50kHz clock speeds of the microprocessor provide sufficiently fast control for the sweep circuitry. The sweep generator employs dual multiplying digital-to-analog convertors (MDACs, Linear Technology LTC1590)  to generate the sawtooth voltage ramps. The sweeps from the individual sources are time-cascaded, generating a MW frequency comb that sweeps different parts of the NV ESR spectrum at once. This allows multiplicative gains in the obtained DNP enhancements. The VCO outputs are power-combined (Minicircuits ZN4PD1-63HP-S+, $p$ = 2.2dBm), passed through a high-isolation switch (Minicircuits ZASWA- 2-50DR+, $p$ = -0.46dBm) and delivered to a low-cost amplifier (Minicircuits ZHL16W-43S+, $p$ = 37.9dBm) that transmits the microwave irradiation to the sample via a stubbed loop antenna (4mm diameter, reflected power $p$ = 36.3dBm, radiation efficiency = 24\%). The radiated MW powers required are extremely low, estimated to be below 1.5W. To estimate the Rabi frequency, we assume an upper limit of microwave output power $P$ = 1.5W at a frequency $\nu$ = 3GHz. Since the circuit is broadband, the magnetic energy is at most $W_B = P/ \nu $ and equal to $w_BV = \frac{B^2}{2\mu_0} \frac{4 \pi}{3}R^3$ where $w_B$ is energy density of the magnetic field and we approximate the volume by a sphere with radius $R$=2mm. Solving for the field, we obtain $B=0.19$mT, and electron Rabi frequency $\Omega_e/2\pi= 430$kHz. To generate the weak field $\Bp$ used for hyperpolarization, as described in \zfr{field}A, we employ a single axis Helmholtz coil (25 turns, 10 layers, 0.8mm diameter) mounted around the sample, generating 14mT fields with $\app$2A of current, and with minimal heating. The coil also helps in the hybrid scenario supplementing detection magnet fringe fields. A set of three small Hall sensors (AsahiKASEI EQ-731L, magnetic sensitivity 65mV/mT) are placed in the polarizing window near the sample, and provide an in-situ measurement of the field. A simple feedback loop can match the field with any desired value.

\I{Materials:} -- The particles employed for hyperpolarization can be HPHT and CVD grown with a high density ($\sim$1ppm) of NV centers. The 100nm particles in \zfr{prototype}E were from Adamas Inc. A variety of features affect the hyperpolarization efficiency including growth and electron irradiation conditions, as well as the methods employed for milling down the particles in size~\cite{Mochalin12}; a systematic study of these aspects will be presented in a forthcoming publication.

\I{Experimental methods:} -- In the experiments of \zfr{prototype}B-E, we employed in the hyperpolarizer device a fiber coupled excitation involving an octagonal arrangement of eight 800mW diode lasers to ensure all particle surfaces were maximally exposed to the illumination.  The field dependence experiments in \zfr{field}A over the wide field range were obtained as a combination of two datasets accessing (1-40mT) and (30-70mT) ranges, and normalizing points in the overlapping regime. From this we obtain the error bar of 11\% indicated in \zfr{field}A. For the latter range, we employed the Synergy DCRO330500  VCOs and AR50S1G6 amplifier combination. For background suppression experiments in \zfr{imaging}, Fmoc-Gly-OH-$\Cs_2$ was successively added in a NMR tube containing 0.5ml DMSO (Fisher Scientific) and 50 particles of 200$\mu$m diamonds at the beginning. After reaching saturation at 1g/ml, the amount of diamonds was halved from step to step. The concentrations in \zfr{imaging}C are displayed normalized to 50 particles. The calculated ratio between the number of $\Cs$ atoms in Fmoc-Gly-OH-$\Cs_2$ and diamond was based on mass considerations. Considering that the Fmoc-Gly-OH-$\Cs_2$ mass $m_F$\:(g), and having molar mass $M_F$ = 299.29g/mol, the molar mass of diamond $M_D$=12.01g/mol, the density of diamond $\rho$ = 3.52 g/cm$^3$, the edge length of the truncated octahedral crystallites $a$ = 87$\pm$3$\mu$m, and the number of diamonds $n$, the ratio of $\Cs$ nuclei outside the diamond to within it can be calculated as $\frac{m_F M_D}{8\sqrt{2}a^3\rho n M_F}\frac{100}{1.1}$ (upper axis in \zfr{imaging}E).

Experiments in \zfr{T1} are carried out by interfacing the hyperpolarizer with a mechanical field cycling instrument constructed over a 7T detection magnet, and consisting of a sensitive conveyor belt actuator stage (Parker HMRB08) with 50$\mu$m precision and 1.5m travel range in the fringe field of the magnet, allowing a rapid($\sim$ 700ms) and wide field sweep range from 10mT-7T.

\bibliography{Biblio}
\bibliographystyle{apsrev4-1}

\pagebreak
\clearpage
\onecolumngrid
\begin{center}
\textbf{\large{\textit{Supplementary Information} \\\smallskip
\bluetitle{Room temperature ``\I{Optical Nanodiamond Hyperpolarizer}'': physics, design and operation}}}\\
\hfill \break
\smallskip
A. Ajoy,$^{1,\textcolor{red}{\ast}}$ R. Nazaryan,$^{1}$ E. Druga,$^{1}$ K. Liu,$^{1}$ A. Aguilar,$^{1}$ B. Han,$^{1}$ M. Gierth,$^{1}$ J. T. Oon,$^{1}$\\ B. Safvati,$^{1}$ R. Tsang,$^{1}$ J. H. Walton,$^{2}$ D. Suter,$^{3}$ C. A. Meriles,$^{4}$ J. A. Reimer,$^{5}$ and A. Pines$^{1}$\\

\smallskip
\emph{${}^{1}$ {\small Department of Chemistry, University of California Berkeley, and Materials Science Division Lawrence Berkeley National Laboratory, Berkeley, California 94720, USA.}}
\emph{${}^{2}$ {\small Nuclear Magnetic Resonance Facility, University of California Davis, Davis, California 95616, USA.}}
\emph{${}^{3}$ {\small Fakultat Physik, Technische Universitat Dortmund, D-44221 Dortmund, Germany.}}
\emph{${}^{4}$ {\small Department of Physics, and CUNY-Graduate Center, CUNY-City College of New York, New York, NY 10031, USA.}}
\emph{${}^{5}$ {\small Department of Chemical and Biomolecular Engineering, and Materials Science Division Lawrence Berkeley National Laboratory University of California, Berkeley, California 94720, USA.}}
\end{center}



\twocolumngrid

\beginsupplement

\begin{figure}[t]
  \centering
  {\includegraphics[width=0.5\textwidth]{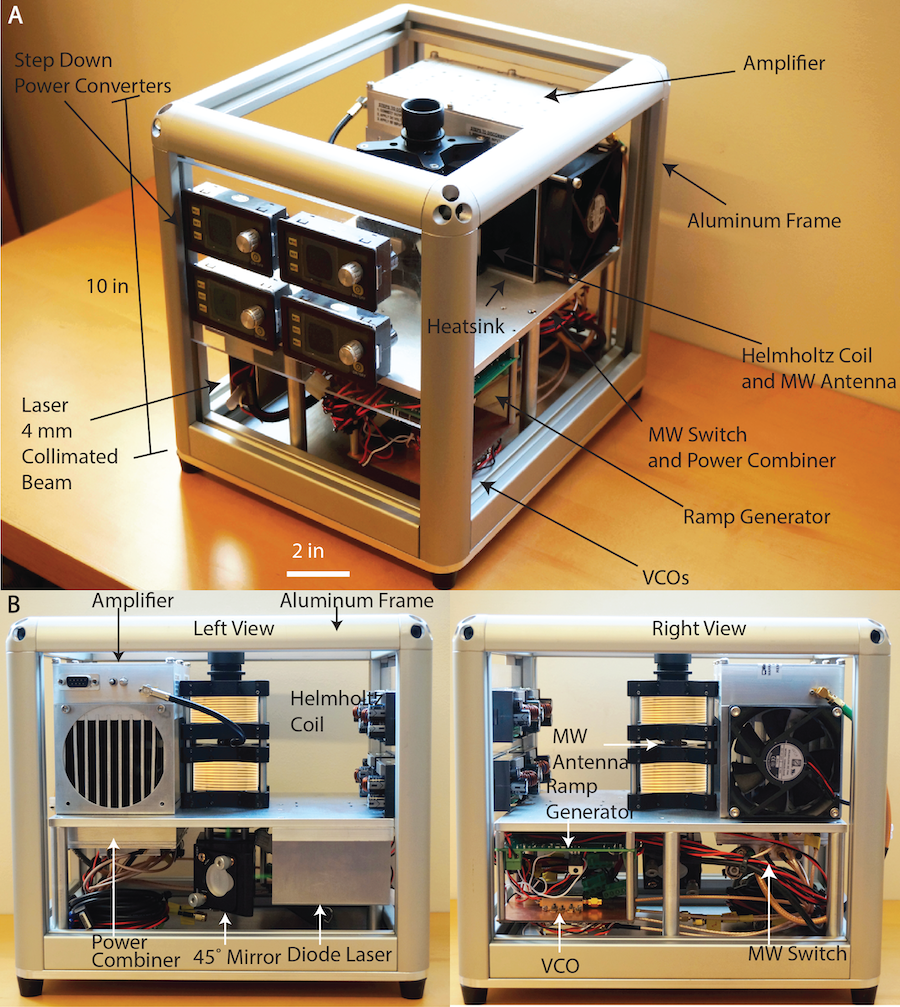}}
  \caption{\T{Ultraportable nanodiamond hyperpolarizer.} (A) \I{Photograph} of device placed on a table-top emphasizing its ultracompact footprint. (B) \I{Side views} showing optical and microwave components that harness the low laser and MW powers required for $\Cs$ hyperpolarization. Polarizing fields ($\Bp\sim$1-45mT) are produced by a simple Helmholtz coil. }
 \zfl{photo-view}	
\end{figure}

\section{Miniature hyperpolarizer construction}
In this section we provide more information about the construction of the hyperpolarizer devices. \zfr{photo-view} shows isometric and side views of device placed on a tabletop. As pointed out in the main paper, all the electronic microwave and optical components are attached to a single aluminum plate (6061 alloy, 0.25")  in the center of the device. The outer frame (8020) acts as scaffolding and provides rigid support to the plate.

To simplify the power handling and delivery, the device consists of a single umbilical cord, interfaced to a rigid 16-pin connector (TME DA-016), that carries power to the various components (Helmholtz coil, amplifiers, VCOs and ramp generators). Step down buck-convertors (DROK) shown on the front panel in \zfr{photo-view} allow the use of a single 28V 10A source to deliver the requisite power to the various devices, obviating the need for separate power supplies for each component. The device side views in panels \zfr{photo-view}B also indicate the relatively simple wiring involved, the power to the devices being carried by means of inexpensive Molex connectors and ribbon cables. The microwave components are interconnected by means of flexible SMA cables (Fairview), while the microwave loop antenna is obtained by stripping and self shorting a semirigid coaxial (MiniCircuits CBL).

\begin{figure}[t]
		 \centering
  {\includegraphics[width=0.42\textwidth]{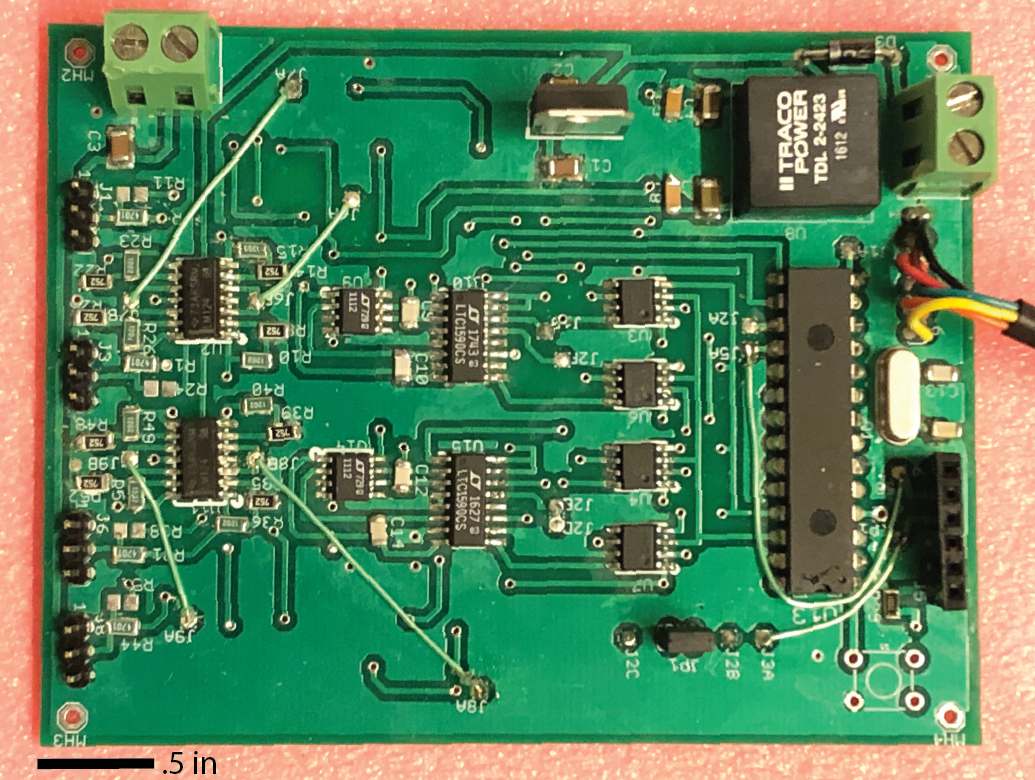}}
  \caption{\T{Homemade quadruple voltage ramp generator} that when interfaced with the VCO sources generate frequency sweeps that drive $\Cs$ nuclear hyperpolarization at low fields. We employ MDACs (LTC1590) to generate sawtooth voltage ramps, the ramp gain and offset values being set by a DAC (MCP4822). All chips are interfaced via a PIC microcontroller running a 50kHz clock. By using a total of two MDAC and DAC chips, we create the quadruple voltage ramp generator that can cascade sweeps from four VCOs to produce the swept MW frequency comb. The entire circuit can be constructed in a miniature footprint.}
\zfl{circuit}	
\end{figure}

\section{Electronics for frequency sweep generation}
At the heart of the DNP process is the use of microwave irradiation that sweeps the NV center ESR spectrum to transfer polarization to the $\Cs$ nuclei. We employ miniature voltage controlled oscillator (VCO) microwave sources for this task. Frequency sweeps are generated by applying sawtooth voltage ramps at the VCO tuning ports. The mean voltage of the ramps are set by the center of the desired sweep band, while the ramp amplitude (gain) is set by sweep bandwidth.
Given the typical frequency-voltage (f-V) characteristics of the sources (Minicircuits ZX95-3800A+), this corresponds to a voltage ramp of 0-20V with the center 10V.

We generate the voltage ramps by means of a home built circuit board employing a PIC microprocessor (PIC30F2020, running at 50kHz), and a multiplying digital to analog convertors (MDAC). The LTC1590 MDACs have 12-bit resolution, and indeed the sawtooth ramps are simple to setup digitally since one just needs to cycle through the binary voltage input values at the desired sweep rates. The ramp gains are configured by a set of independent serial DACs (MCP4822), that are also controlled by the PIC. Since the MDACs produce AC sawtooth ramps, they have to be DC offset to the center of the sweep band. This is achieved by using another DAC, and finally summing them via an operational amplifier (LM124) noninverting adder and gain stage. Note that both the MDACs, DACs and opamps we employ are all dual chip packages and thus allow the use of fewer overall components.

Importantly this also enables the simple cascading of multiple VCO sources to gain in DNP enhancements. Effectively, the $N$ microwave sources can be combined to generate swept MW frequency combs, that can sweep different parts of the ESR spectrum simultaneously and obtain multiplicative DNP gains that scale linearly with $N$. The use of two MDACs, DACs, and four opamps allow one to cascade four VCO sources, which we have found to be optimal given the relatively narrow frequency bandwidths at low fields.

\begin{figure}[t]
  \centering
  {\includegraphics[width=0.5\textwidth]{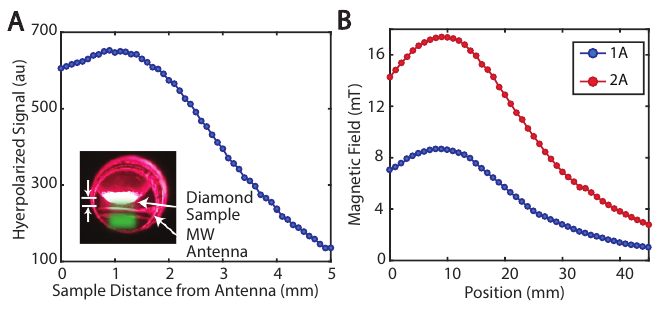}}
  \caption{\T{Experimental characterization of MW and field inhomogeneity.} (A) \I{MW inhomogeneity.} Using a thin single crystal sample (3\% $\Cs$ enriched) with thickness $\app$250$\mu$m we map the effect of MW inhomogeneity in our DNP experiments via the obtained $\Cs$ hyperpolarization enhancement. We use a 4mm MW loop antenna in these experiments, and the results indicate that a volume of over 10mm$^3$ can be hyperpolarized in under 60s of pumping \I{Inset:} schematic of experiment where sample height above the loop antenna is varied in 100$\mu$m increments. (B) \I{Field inhomogeneity.} We measure (via a Hall probe) the polarizing magnetic field produced by the Helmholtz coils as a function of sample height for 1A and 2A applied current. Panels indicates that $\Cs$ hyperpolarization can function even under highly inhomogeneous MW settings and field environments. }
 \zfl{height}	
\end{figure}

\section{Device performance under field and MW inhomogeneity}
We now detail experiments that characterize the performance of the hyperpolarizer in the presence of microwave irradiation for polarizing field inhomogeneities. In \zfr{height}A, we employ a 3\% enriched $\Cs$ single-crystal sample of thickness $\app$250$\mu$m mounted flat in an 8mm NMR tube (see inset), and a 4mm microwave loop antenna interfaced to the 16W MW amplifier for DNP excitation. The sample is positioned \I{above} the antenna and subsequently moved in incremental 100$\mu$m steps and the hyperpolarization enhancements are recorded at each sample height. Experiments are performed in the $\Bpol=$20mT fringe field of a 7T magnet, which has a very slowly varying field gradient and hence the obtained enhancements report the device performance under varying microwave powers. As is evident (\zfr{height}A) far away from the antenna the loss in enhancement is approximately linear with distance, which is on account of the microwave power being too low to efficiently excite DNP. On the other hand we notice a relatively homogeneous region approximately constrained by the radius of the loop, where the enhancements remain fairly constant, indicating the ability to efficiently excite DNP in a $\sim$16mm$^3$ hemispherical volume. Considering the density of diamond ($ \rho = 3.51$mg/mm$^3$), this corresponds to a total mass of 50mg, and a hyperpolarization throughput of $\sim$50mg/min since DNP buildup saturates in about a minute (Fig. 1B of main paper).

In \zfr{height}B we study the homogenity of the polarizing field $\Bpol$ produced by the Helmholtz coil (shown for two current values, 1A and 2A, corresponding respectively to coil field values of 70.2G and 142.7G) here defined as the percent difference of the magnetic field from the central field intensity. Measurements are obtained by mounting a sensitive Hall probe (Lakeshore HMMA-2504-VR-10) on a mechanical stage and measuring the polarizing field as a function of height. Given that the heights of the packed diamond powder typically used in device operation (2-3mm placed in the center of the Helmholtz pair), limited in part by optical penetration depth, this corresponds to a magnetic field homogenity of $<$10\% for a separation distance of 3mm at an operating current of 2A. Note that DNP on powders is relatively immune to inhomogeneity in the polarizing field, since the NV centers electronic spectrum is already orientationally broadened by several hundred MHz. Inhomogeneity only adds a slight additional broadening, and does not significantly affect the obtained DNP enhancements.

\begin{figure}[t]
  \centering
  {\includegraphics[width=0.4\textwidth]{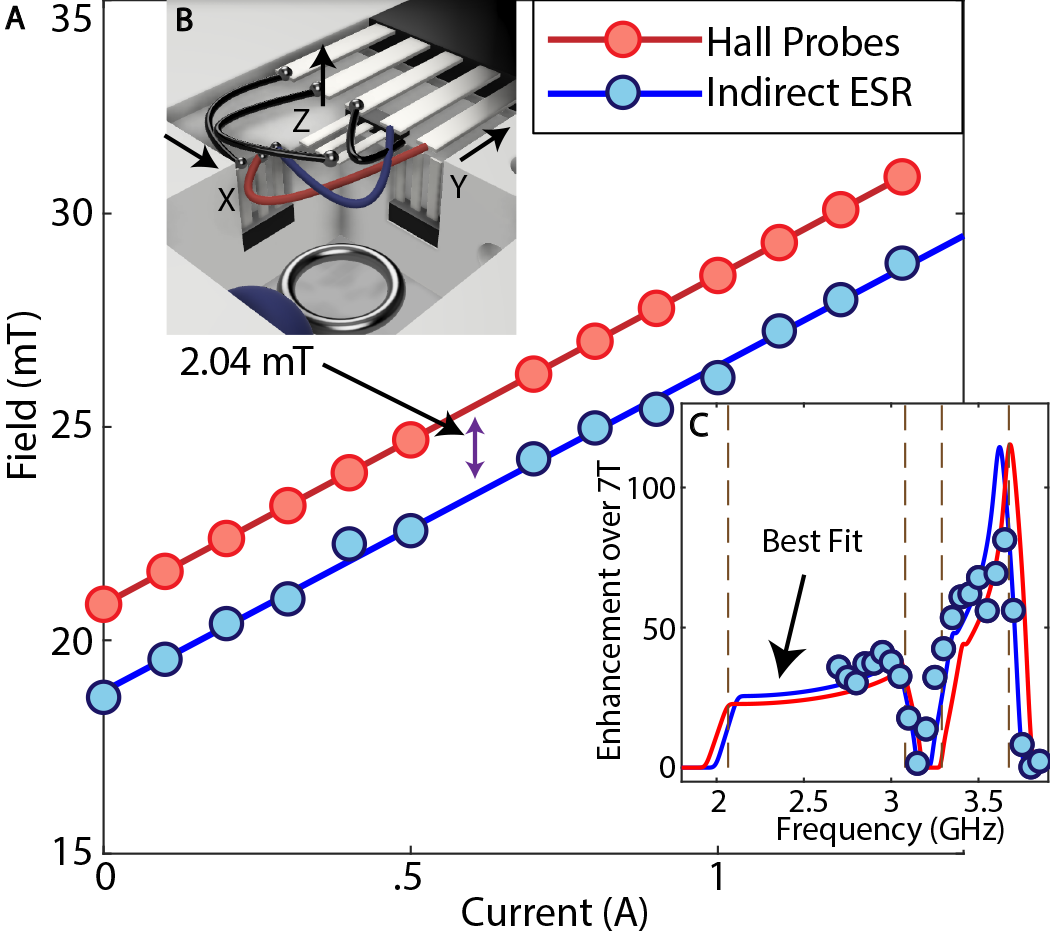}}
  \caption{\T{Experimental characterization of Hall probes for field estimation.} (A) Red (blue) lines indicate the measured (estimated) fields from the Hall probes (indirectly from ESR measurements of the NV power pattern). Field is produced by the Helmholtz coils in the 20mT fringing field of a 7T NMR magnet. (B) Panel highlights the embedded tri-orthogonal Hall probes allowing one to measure the magnetic field in the vicinity of the sample being hyperpolarized. (C) Using 100MHz sweep windows we map out experimentally the NV ESR spectrum (points). Blue line indicates best fit to the simulated ESR spectrum, while the red line is the predicted spectrum using the measured values from the Hall probes alone. We therefore slightly overestimate the field using the Hall probes, by a relatively small amount $\app$2.04mT, which we ascribe to being because the $\bz$-axis Hall probe is $\app$1mm above the microwave loop antenna.}
 \zfl{hall}	
\end{figure}

\section{Hall probes for field estimation}
We now provide more details of the embedded Hall sensors (AsahiKASEI EQ-731L, magnetic sensitivity$\sim$65mV/mT) that allow the in-situ measurement of magnetic field used for hyperpolarization experiments (Fig. 3A of main paper). The probes enable high flexibility in the operation and installation of the device in the fringe fields of NMR detection magnets, since an evaluation of the $\Bp$ field allows one to precisely estimate the frequency range to sweep over for optimal DNP enhancements (Fig. 3 of main paper), as well as allowing to supplement the fringe field with an applied current in the Helmholtz coils.

The Hall probes are mounted in a tri-orthogonal configuration in the vicinity of the microwave loop antenna (see \zfr{hall}B for a zoomed view). This allows the measurement of the vector magnetic field $\Bpol$ seen by the NV electrons during DNP. In order to characterize the accuracy of the measured field value, \zfr{hall}C shows the comparison between the directly measured field, and the field estimated by fits of the indirectly obtained NV center powder patterns via hyperpolarized $\Cs$ NMR by performing DNP experiments on 100MHz windows that are swept across in frequency space (see Fig. 2B of main paper). We obtain a close and consistent agreement between the two methods (see \zfr{hall}C), the Hall measurement being offset from the exact field felt by NV centers by $\app$2mT. We ascribe this slight overestimation to be because the Hall sensor in the $\bz$ direction is about 1mm away from the loop antenna (see \zfr{hall}C). In summary the use of these miniature low-cost Hall sensor chips enable the versatile operation of the hyperpolarizer device in a variety of field environments.

 \begin{figure}[t]
 		 \centering
   {\includegraphics[width=0.33\textwidth]{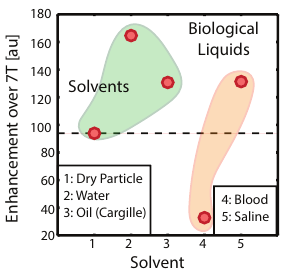}}
   \caption{\T{DNP of particles in solution.} 200$\mu$m natural abundance particles immersed in solvents and biologically relevant liquids.  Variations arise from differences in optical penetration. Shaded regions are guides for the eye.   Since the DNP occurs at room temperature and with modest laser and MW powers, large hyperpolarization enhancements can be gained even when the particles are placed in optically nontransparent liquids such as blood. }
   \zfl{solvents}	
 \end{figure}
 
 \begin{figure*}[t]
  \centering
  {\includegraphics[width=0.98\textwidth]{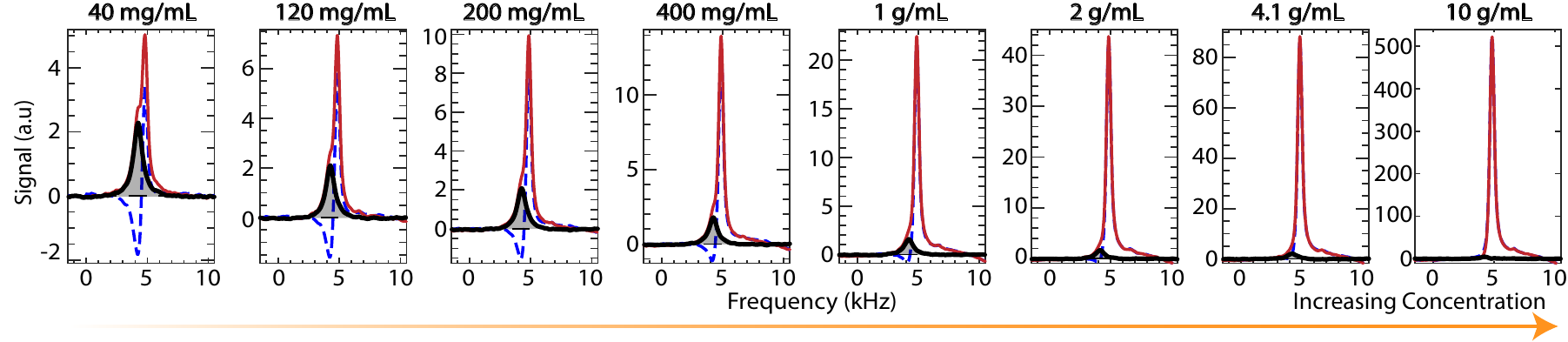}}
   \caption{\T{Background suppression experiments.} Full data set of experiments depicted in Fig. 6C of the main paper. Here the red solid (blue dashed) lines indicate the $\Cs$ NMR spectrum obtained with low to high (high to low) frequency sweeps. Black lines and shaded regions are extracted the $\Cs$ diamond spectra obtained by subtracting the results. The panels show increasing concentrations of Fmoc-Gly-OH-$\Cs_2$ background signal, normalized here for 50 diamond particles (see Methods). In the last panel, also depicted in Fig. 6D of the main paper, the peak suppression of the background signal exceeds two orders of magnitude.}
 \zfl{fmoc-full}	
\end{figure*}

\section{DNP of particles in solution}
Our hyperpolarizer is fully operational at room temperature and employs very modest amount of laser and MW powers. This allows the DNP process in diamond performed at ambient conditions with the particles dry as well as in solution. In \zfr{solvents} we study this in detail for particles immersed in a wide range of solutions, including common solvents and biological fluids. We find that high enhancements can be preserved. It is noteworthy that hyperpolarization was possible rather efficiently even for particles that were immersed in blood which is viscous and less optically clear (\zfr{solvents}). Overall this illustrates the advantages of room temperature DNP at low magnetic fields and employing low optical and microwave powers.

This also enables the experiments in Fig. 6C of the main paper where we perform DNP of particles in a solution of $\Cs$ enriched the liquid, having a chemical shift very close to the diamond peak, and showed how by successive experiments with alternate hyperpolarization signs one could suppress very effectively this $\Cs$ background. For completeness in \zfr{fmoc-full}, we present the complete data set that was used in Fig. 6E of the main paper. In the last panel, (also shown in Fig. 6D), we obtain a $\Cs$ background suppression exceeding two orders of magnitude.

\section{Measurement of field dependent $\Cs$ lifetimes}
We outline herein the measurement strategy and error estimation in experiments presented in Fig. 4 of the main paper studying the field dependence of $^{13}$C nuclear lifetimes by interfacing the portable hyperpolarizer with a home built field cycling instrument (see Methods). The instrument consists of a high precision (50$\mu$m) conveyor belt actuator stage in the fringe field of a 7T magnet, allowing a fine step field sweep over a large range from $\sim$10mT-7T. We are able to translate the sample between different fields precisely with sub-second switching speeds (measured to be $648\pm4$ms for full 20mT-7T travel), and with little loss in enhancement. We measure a striking field dependence (see Fig. 4), the nuclear lifetimes falling rapidly below a field of $\approx$57mT and reaching remarkably high values $>$6min beyond 100mT.

In order to hasten measurement times, and to obtain a map of nuclear $T_1$ at a large number (55) of field points, we use an accelerated measurement strategy detailed below. We first measure the full $T_1$ relaxation decays for a subset of points, $\sim$20 in number, uniformly sampled from low to high field. To a very good approximation we find that the relaxation decays are mono-exponential (see inset of Fig. 4C of main paper). Next we measure for all 55 field values an accelerated 1D relaxation measurement where we measure the signal decay after a fixed waiting time (60s) after hyperpolarization and subsequent rapid transfer to the field of interest. Due to the reduced dimensionality of this measurement, it is possible to rapidly measure a large number of field points. We now map the obtained field-dependent enhancement decay values from this 1D data set to nuclear $T_1$ values assuming mono-exponential decays. 

This is done by first calculating the enhancement at zero time $\epsilon(t=0)$ at fields with both enhancement decay and 1D measurements. Using the $T_1\pm \sigma_{T_1}$ estimates from our fit, we solve for this initial enhancement $\epsilon(0) = \epsilon(60)e^{\frac{60}{T_1}}$ assuming mono-exponential decay, with corresponding error
$\sigma_{\epsilon(0)} = \frac{60 \cdot \epsilon(60)\cdot \sigma_{T_1} e^{60/T_1} }{T_1^2}$.
We average our calculations of $\epsilon(0)$ over measurements at high field ($>$500mT), as fast relaxation times at low fields make it challenging to obtain accurate estimations of $\epsilon(0)$. During calculation of the mean, we assigned weights $w = \sigma_{\epsilon(0)}^{-2}$ for each measurement. Thus for N values of $\epsilon(0)$, our resulting mean $\overline{\epsilon}(0)$ has error $\sigma_{\overline{\epsilon}(0)} = \left(\sqrt{\sum_i^N w_i}\right)^{-1}$. We then map our 60s enhancement measurements $\epsilon(60)$ to $T_1 = 60/\log\left(\frac{\overline{\epsilon}(0)}{\epsilon(60)}\right)$, with associated error $\sigma_{T_1} = \frac{60 \cdot \sigma_{\overline{\epsilon}(0)}}{\overline{\epsilon}(0)}/\left[\log \left(\frac{\overline{\epsilon}(0)}{\epsilon(60)}\right)\right]^2$. This is the data plotted in Fig. 4 of the main paper.

In order to fit the obtained relaxation profiles, we fit a Lorentzian decay to the relaxation rate at low fields and assume a constant offset to the rates at high field. Its functional form with respect to field $B$ is $R_1(B)=\fr{1}{T_1(B)}=\frac{2A}{\pi} \frac{W}{4 B^2 + W^2} + c$, with fitting parameters $A, W, c$ describing the amplitude, width and vertical offset of the Lorentzian respectively. 

We have found a qualitatively similar field profile for all 1\% $^{13}$C (natural abundance) including for single crystals and powders (5-200$\mu$m) both CVD as well as HPHT manufactured. A more detailed study of the factors that cause this dramatic field dependence, and accompanying ESR measurements of these diamond samples, will be presented in a forthcoming publication.

\begin{figure}
  \centering
  {\includegraphics[width=0.4\textwidth]{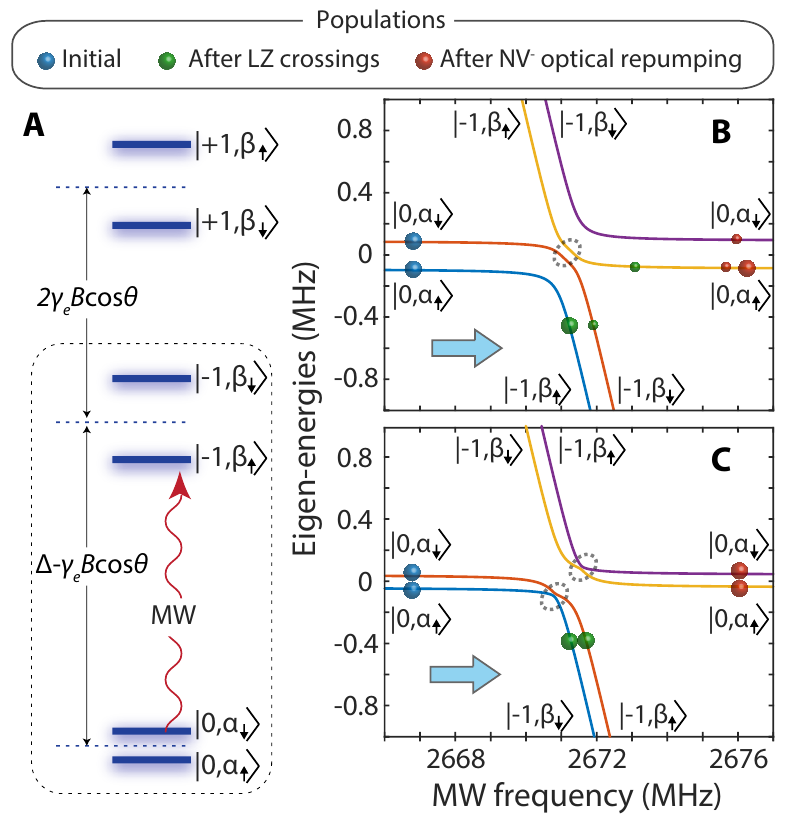}}
  \caption{
\textbf{Mechanism of polarization transfer.} (A) Energy levels of an NV electron spin hyperfine-coupled to a $\Cs$ nuclear spin. Quantum numbers in all kets refer to electron and nuclear spins, in that order; the notation for the nuclear spin states highlights the manifold-dependent quantization axis, in general different from the magnetic field direction. (B) Calculated energy diagram in the rotating frame corresponding to the $m_S=0\leftrightarrow m_S=-1$ subset of transitions (dashed rectangle in (A)) assuming a hyperfine coupling $A_{zz}=+0.5$ MHz. (C) Same as in (B) but for $A_{zz}=-0.5$ MHz. In (B) and (C) we assume $B$=10 mT, $\theta$=45 deg., and use a transverse hyperfine constant $A_{zx}=0.3|A_{zz}|$. Colored solid circles denote populations at different stages during a sweep in the direction of the arrow, and faint dashed circles indicate the narrower avoided crossings where population transfer takes place. }
\zfl{mechanism}
\end{figure}

\section{Hyperpolarization Mechanism}
We now briefly describe the low field DNP mechanism that governs the polarization transfer in our experiments. For more details, and experimental characterization of the mechanism, we point the reader to Ref. \cite{Ajoy17, Zangara18}. Consider for simplicity a NV center coupled to a single $\Cs$ nuclear spin. The Hamiltonian of the system is,
\beq\label{eq:1}
\begin{aligned}
	\mH = {} & \xD S_{z}^2 - \xg_e \vec{B} \cdot \vec{S} - \xg_n \vec{B} \cdot \vec{I} + A_{zz}S_{z}I_{z}\\
	& + A_{yy}S_{y}I_{y} + A_{xx}S_{x}I_{x} + A_{xz}S_{x}I_{z} + A_{zx}S_{z}I_{x}
\end{aligned}
\eeq
where $\vec{S}$ and $\vec{I}$ respectively denote the NV and $\Cs$ vector spin operators, and $\vec{B}$ is the magnetic field (10-30 mT) at angle $\theta$ ($\phi$) to the NV axis. Within the $m_s=\pm1$ states, the hyperfine coupling produces a $\Cs$ splitting,
\beq\label{eq:2}
	\xo_{C}^{(\pm1)} = \sqrt{(A_{zz}\mp\xg_{n}B\cos\xt)^2 + A_{zx}^2}
\eeq
For the $m_s=0$ manifold, second-order perturbation theory leads to the approximate formula~\cite{Alvarez15},
\beq\label{eq:3}
\begin{aligned}
	\wt{\xo}_L \approx {} & \xg_{n}B \\
	& + 2\Big(\frac{\xg_{e}B}{\xD}\Big) \sin\xt \big(\sqrt{A_{xx}^2 + A_{zx}^2} \cos^2\phi  + A_{yy}\sin^2\phi \big)
\end{aligned}
\eeq 
From Eqs. \ref{eq:2} and \ref{eq:3} we conclude that each manifold (including the $m_s=0$ manifold) has its own, distinct quantization axis which might be different from the direction of the applied magnetic field. In particular, the second term in Eq. \ref{eq:3} can be dominant for hyperfine couplings as low as 1 MHz (corresponding to nuclei beyond the first two shells around the NV) if $\theta$ is sufficiently large, implying that, in general, $\Cs$ spins coupled to NVs misaligned with the external magnetic field experience a large frequency mismatch with bulk carbons, even if optical excitation makes $m_s=0$ the preferred NV spin state.

Assuming fields in the range 10-30 mT, it follows that $\Cs$ spins moderately coupled to the NV (300 kHz $\lesssim |A_{zz}|\lesssim$ 1 MHz) are dominant in the hyperpolarization process because they more easily spin diffuse into the bulk and contribute most strongly to the observed NMR signal at 7T. For sweep rates near the optimum ($\sim$ 40 MHz/ms), the time necessary to traverse the set of transitions connecting $m_s=0$ with either the $m_s=-1$ or $m_s=+1$ manifolds is relatively short \big($\lesssim$ 30 $\mu$s for weakly coupled carbons\big) meaning that optical repolarization of the NV preferentially takes place during the longer intervals separating two consecutive sweeps, as modeled in \zfr{mechanism}. 

Nuclear spin polarization can be understood as arising from the Landau-Zener crossings in \zfr{mechanism}. Efficient polarization transfer takes place when the narrower LZ crossings connect branches with different electron and nuclear spin quantum numbers, precisely the case in the $m_s=0\leftrightarrow m_s=-1$ ($m_s=0\leftrightarrow m_s=+1$) subset of transitions when the hyperfine coupling is positive (negative). When probing ensembles, both sets of transitions behave in the same way, i.e., $\Cs$ spins polarize positive in one direction, negative in the other. A more detailed exposition of the hyperpolarization mechanism and simulations are presented in Ref. \cite{Ajoy17}.

\end{document}